\documentclass[sigconf]{acmart}
\usepackage{xspace}
\usepackage{xcolor}
\usepackage{color}
\usepackage{colortbl}
\usepackage{makecell}
\usepackage{multirow}
\usepackage{geometry}
\usepackage{hhline}
\usepackage{marginnote}
\usepackage{hyperref}
\usepackage[normalem]{ulem}
\usepackage{pifont}
\usepackage{enumitem}

\usepackage[lined,boxed,vlined,ruled,linesnumbered]{algorithm2e}

\usepackage{framed}
\definecolor{shadecolor}{rgb}{0.92,0.92,0.92}

\newcommand{\zxh}[1]{\textcolor{blue}{#1}}
\newcommand{\xbr}[1]{\textcolor{teal}{\textbf{BR:} #1}}

\usepackage[switch]{lineno}

\makeatletter

\newcommand{\llabel}[1]{}

\newcommand{\nlabel}[1]{}

\makeatother

\newcommand{\gray}[1]{\textcolor{gray}{#1}}

\setcopyright{none}
\renewcommand\footnotetextcopyrightpermission[1]{} 
\makeatother

\usepackage{tcolorbox}
\tcbuselibrary{breakable}

\usepackage{marginnote}

\newtcolorbox{MetaBox}{
  breakable,
  colback=green!5,      
  colframe=green!30,    
  boxrule=0.5pt,       
  arc=2pt,             
  left=5pt, right=5pt, top=5pt, bottom=5pt, 
  fontupper=\small 
}

\newtcolorbox{R1Box}{
  breakable,
  colback=blue!5,      
  colframe=blue!30,    
  boxrule=0.5pt,       
  arc=2pt,             
  left=5pt, right=5pt, top=5pt, bottom=5pt, 
  fontupper=\small 
}

\newtcolorbox{R3Box}{
  breakable,
  colback=orange!5,      
  colframe=orange!30,    
  boxrule=0.5pt,       
  arc=2pt,             
  left=5pt, right=5pt, top=5pt, bottom=5pt, 
  fontupper=\small 
}

\newtcolorbox{R4Box}{
  breakable,
  colback=red!5,      
  colframe=red!30,    
  boxrule=0.5pt,       
  arc=2pt,             
  left=5pt, right=5pt, top=5pt, bottom=5pt, 
  fontupper=\small 
}

\newcommand{\MetaColor}[1]{#1}
\newcommand{\RaColor}[1]{#1}
\newcommand{\RbColor}[1]{#1}
\newcommand{\RcColor}[1]{#1}
\renewcommand\marginpar[2][]{}

\colorlet{BLUE}{blue}
\colorlet{ORANGE}{orange!80!black}
\colorlet{RED}{red}

\newcommand\vldbdoi{XX.XX/XXX.XX}
\newcommand\vldbpages{XXX-XXX}
\newcommand\vldbvolume{14}
\newcommand\vldbissue{1}
\newcommand\vldbyear{2020}
\newcommand\vldbauthors{\authors}
\newcommand\vldbtitle{\shorttitle} 
\newcommand\vldbavailabilityurl{URL_TO_YOUR_ARTIFACTS}
\newcommand\vldbpagestyle{plain} 

\newcommand{\hi}[1]{\vspace{.25em} \noindent {\bf #1} }
\newcommand{\llm}{\textsc{LLM}\xspace}

\newcommand{\bfit}[1]{\textbf{\textit{#1}}}
\newcommand{\oursys}{\texttt{MoDora}\xspace}
\newcommand{\ourbench}{\texttt{MMDA}\xspace}

\newcommand{\modocanalysis}{{MoDocAnalysis}\xspace}

\usepackage{bm}

\theoremstyle{definition}
\newtheorem{definition}{Definition}[section]

\makeatletter
\newcommand{\removelatexerror}{\let\@latex@error\@gobble}
\makeatother

\begin{document}
\title{DocScope: LLM-Powered Multi-Modal Document Analysis}
\title{\oursys: An LLM-Powered Lens for Multi-Modal Document Analysis}
\title{MoDora: A Tree-Based Framework for Semi-Structured\\ Document Analysis with LLMs}
\title{MoDora: Tree-Based Semi-Structured Document Analysis System}

\pagenumbering{arabic}

\setcounter{page}{1}
\setcounter{figure}{0}
\setcounter{table}{0}

\author{Bangrui Xu}
\affiliation{%
  \institution{Shanghai Jiao Tong University}
}
\email{dreameternal@sjtu.edu.cn}

\author{Qihang Yao}
\affiliation{%
  \institution{Shanghai Jiao Tong University}
}
\email{yaoqihang@sjtu.edu.cn}

\author{Zirui Tang}
\affiliation{%
  \institution{Shanghai Jiao Tong University}
}
\email{tangzirui@sjtu.edu.cn}

\author{Xuanhe Zhou}\authornote{Xuanhe Zhou and Yeye He are the corresponding authors.}
\affiliation{%
  \institution{Shanghai Jiao Tong University}
}
\email{zhouxuanhe@sjtu.edu.cn}

\author{Yeye He*}
\affiliation{%
  \institution{Microsoft Research}
}
\email{yeyehe@microsoft.com}

\author{Shihan Yu}
\affiliation{%
  \institution{Beihang University}
}
\email{yushihan070611@qq.com}

\author{Qianqian Xu}
\affiliation{%
  \institution{Beihang University}
}
\email{24421025@buaa.edu.cn}

\author{Bin Wang}
\affiliation{%
  \institution{Shanghai AI Lab}
}
\email{wangbin@pjlab.org.cn}

\author{Guoliang Li}
\affiliation{%
  \institution{Tsinghua University}
}
\email{liguoliang@tsinghua.edu.cn}

\author{Conghui He}
\affiliation{%
  \institution{Shanghai AI Lab}
}
\email{heconghui@pjlab.org.cn}

\author{Fan Wu}
\affiliation{%
  \institution{Shanghai Jiao Tong University}
}
\email{fwu@cs.sjtu.edu.cn}
\begin{abstract}
Semi-structured documents integrate diverse interleaved data elements (e.g., tables, charts, hierarchical paragraphs) arranged in various and often irregular layouts. These documents are widely observed across domains and account for a large portion of real-world data. However, existing methods struggle to support natural language question answering over these documents due to three main technical challenges: (1) The elements extracted by techniques like OCR are often fragmented and stripped of their original semantic context, making them inadequate for analysis. (2) Existing approaches lack effective representations to capture hierarchical structures within documents (e.g., associating tables with nested chapter titles) and to preserve layout-specific distinctions (e.g., differentiating sidebars from main content). {(3)} Answering questions often requires retrieving and aligning relevant information scattered across multiple regions or pages, such as linking a descriptive paragraph to table cells located elsewhere in the document.

To address these issues, we propose \oursys, an LLM-powered system for semi-structured document analysis. First, we adopt a local-alignment aggregation strategy to convert OCR-parsed elements into layout-aware components, and conduct type-specific information extraction for components with hierarchical titles or non-text elements. Second, we design the {Component-Correlation Tree} (CCTree) to hierarchically organize components, explicitly modeling inter-component relations and layout distinctions through a bottom-up cascade summarization process. Finally, we propose a question-type-aware retrieval strategy that supports (1) layout-based grid partitioning for location-based retrieval and (2) LLM-guided pruning for semantic-based retrieval. Experiments show \oursys outperforms baselines by 5.97\%-61.07\% in accuracy. The code is at {\it \textcolor{blue}{\url{https://github.com/weAIDB/MoDora}}}.

\end{abstract}

\maketitle
\renewcommand{\shortauthors}{Bangrui Xu et al.}

\begin{sloppypar}




\section{Introduction}
\label{sec:intro}


\begin{figure}[!t]
    \centering
    \includegraphics[width=1\linewidth]{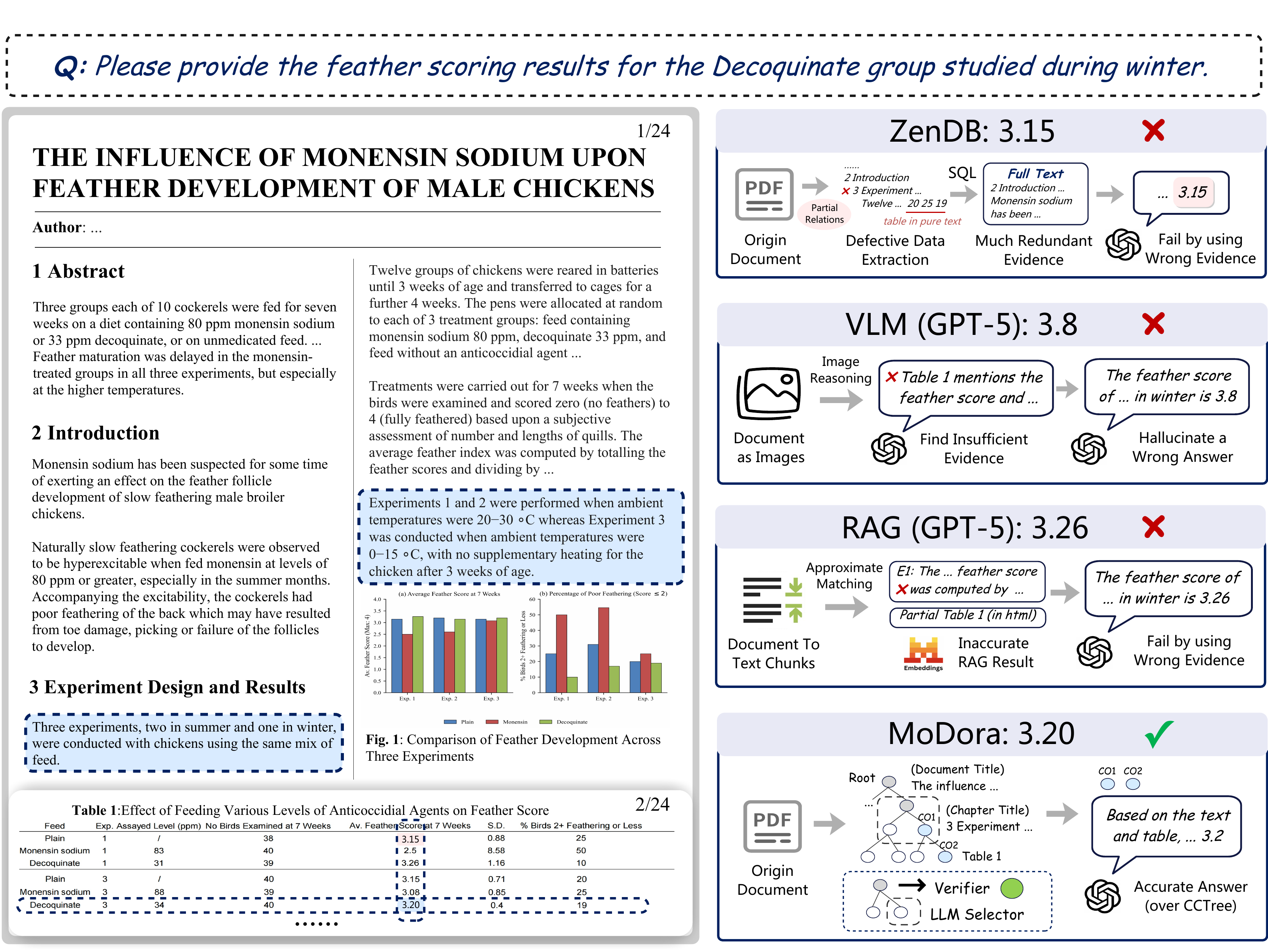}
    \vspace{-1.75em}
    \caption{Example Semi-Structured Document Analysis.}
    \label{fig:intro}
\end{figure}

Semi-structured documents contain a mixture of diverse elements, such as tables, charts, free-form text, arranged in complex and often irregular layouts. These documents are widely observed in real-world scenarios, including scientific reports, financial statements, and technical manuals~\cite{arora2023language}.


\begin{example}
\label{ex:1}
{\it Figure~\ref{fig:intro} presents a typical semi-structured scientific report containing components (COs) like (1) a data table of experimental results, (2) a chart visualizing trends from the table, and (3) accompanying text describing the setup and findings. While the table and chart offer structured insights, accurate question answering requires holistic interpretation across all components. Additionally, the report's hierarchical titles reflect the nested structures commonly found in semi-structured formats like HTML and JSON~\cite{twix}.}
\end{example}


Based on our analysis of over one million real-world documents\footnote{Collected from representative sources like Scihub~\cite{scihub} and PDF drive~\cite{pdfdrive} (\href{https://anonymous.4open.science/r/MoDora-3415/README.md}{\gray{[\underline{artifact}]}}).}, we find that over 77\% of the data contain at least one table, chart, or paragraph title. Among these, 61\% include at least one table and 40\% contain at least one chart, demonstrating the ubiquity of documents with structured elements interleaved with free-form text.



\emph{Semi-structured document analysis} aims to support natural language (NL) question answering over such documents, which is common and practically valuable. For the example in Figure~\ref{fig:intro}, to answer the question ($\bm{Q}$) about the feather state of specific group in winter, a human analyst needs to first infer from paragraphs on page 1 that Experiment 3 was conducted in winter, and then locates the corresponding feather score from the table on page 2.



Existing relevant methods can be broadly categorized into five classes, each exhibiting critical limitations in semi-structured document analysis~\cite{tang2023unifying, openai2025gpt5, hu2024mplug, cho2024m3docrag, chen2024sv, sun2025quest, zendb, hegde2023analyzing, liu2025palimpzest, arora2023language, huang2022layoutlmv3, appalaraju2021docformer, sun2025docagent, yang2025superrag, saad2024pdftriage}: (1) Content extraction methods such as QUEST~\cite{sun2025quest} and EVAPORATE~\cite{arora2023language} extract information (e.g., attribute–value pairs) and store it in relational tables for structured retrieval. While efficient for text content extraction, they discard document structure information, making it difficult to answer questions that depend on layout cues. 
\marginpar[]{\textbf{\MetaColor{Meta.O1}}}
\marginpar[]{\textbf{\RbColor{R2.O1-(1),(2)}}}{\RbColor{\llabel{more_work3}(2) Structure extraction methods such as ZenDB~\cite{zendb}, DocAgent~\cite{sun2025docagent}, SuperRAG~\cite{yang2025superrag}, and PDFTriage~\cite{saad2024pdftriage} can extract hierarchical document structures (e.g., chapters and page mappings). However, they suffer from two main limitations. First, these methods typically represent documents at the page level, lacking fine-grained modeling. For instance, DocAgent does not explicitly model region-level coordinates within pages, which hinders precise location-based retrieval. Similarly,  ZenDB fails to capture the structural and semantic relations between non-textual elements (e.g., tables, charts) and surrounding text, resulting in fragmented tree representations that limit downstream analysis. 
Second, the structural modeling methods lack robustness. For instance, ZenDB relies on visual feature clustering and LLM to detect titles and their body text. However, such clustering is error-prone, especially for complex document layouts, and can cause incorrect or flattened hierarchies.
}} (3) Programmatic methods like Palimpzest~\cite{liu2025palimpzest} require manually written domain-specific language (DSL) programs to operate over documents. This limits applicability in natural-language QA scenarios where no programs are provided. 
\marginpar[\textbf{\MetaColor{Meta.O1}}]{}
\marginpar[\textbf{\RaColor{R1.O3}}]{}{\RaColor{\llabel{more_work1} 
(4) End-to-End models including pre-trained models (e.g., LayoutLMv3~\cite{huang2022layoutlmv3}, DocFormer~\cite{appalaraju2021docformer}, UDOP~\cite{tang2023unifying}) and Multimodal Large Language Models (MLLMs), such as DocOwl2~\cite{hu2024mplug} and GPT-5~\cite{openai2025gpt5}, process documents as images and answer in an end-to-end manner. However, these models often miss fine-grained textual details and above layout-specific relations, causing incomplete evidence aggregation.}} 
(5) Retrieval-Augmented Generation (RAG) methods~\cite{cho2024m3docrag,chen2024sv,hegde2023analyzing,yang2025superrag} retrieve content through embedding and lack structural alignment, resulting in missing or misaligned evidence during multi-element reasoning. TextRAG retrieves relevant semantic pieces from chunks of OCR output~\cite{hegde2023analyzing}, while SV-RAG~\cite{chen2024sv} and M3DocRAG~\cite{cho2024m3docrag} retrieve relevant page images.

We find these existing methods all fall short in understanding semi-structured documents, like illustrated in the example below.

\begin{example} {\it Continue with Example~\ref{ex:1}, when given a document and a question like the one shown in Figure~\ref{fig:intro}, we observe that ZenDB fails to answer primarily because it breaks the structure of Table~1 by flattening it into plain text under Chapter~3. This disruption makes it difficult to align the \underline{text paragraph} mentioning ``winter'' with the relevant \underline{table region} (\textit{Decoquinate, Experiment 3}) needed to identify the correct answer (3.20). In addition, ZenDB misidentifies the document hierarchy (e.g., nesting Chapter~3 under Chapter~2) and retrieves overly redundant content (e.g., the entire document), further impeding it to give the correct answer. Similarly, VLM (GPT-5) hallucinates an answer (3.8) by only analyzing the table image, while overlooking the surrounding paragraph that indicates \textit{Experiment 3} was conducted in winter. {RAG (GPT-5)} retrieves the sentence describing how the average feather score was computed, but fails to align the retrieved Table 1 (in HTML-rendered chunk) and produces an incorrect value (3.26). In contrast, we aim to aggregate elements into semantically meaningful units (e.g., tables with titles and positions) and explicitly capture their relationships and layout distinctions, so as to enable joint reasoning (e.g., the target paragraph and table row) and give correct answer. 
}
\end{example}

\hi{Challenges.} The technical challenges of this work are threefold.

First, the elements extracted by OCR (e.g., paragraphs, tables, charts) are often fragmented and detached from their semantic context, making it difficult to interpret them in isolation. This hinders fine-grained document structure understanding (i.e., component level), such as associating a chart with its title or aligning a paragraph with a nearby table (\textbf{C1}). 

Second, existing approaches lack effective representations to capture hierarchical structures (e.g., tables under nested chapter titles) and to maintain layout-specific distinctions (e.g., differentiating sidebars from main content). For instance, while scientific reports follow a nested chapter structure with interleaved visual elements, newspapers often contain parallel articles. Current methods either treat documents as flat sequences~\cite{sun2025quest,hegde2023analyzing} or process them page-wise~\cite{cho2024m3docrag,chen2024sv}, thus failing to model global structure (document level) and cross-component relationships (\textbf{C2}). 

Third, answering complex questions requires retrieving and aligning evidence scattered across multiple regions or pages. For instance, in Figure~\ref{fig:intro}, the answer requires (1) identifying an experiment ID from the question, (2) linking it to a season mentioned in a paragraph, and (3) locating the corresponding value in table within another page. Moreover, retrieval must support both semantic references (e.g., summary of nested chapters) and spatial ones like \textit{``top of page 1''} (\textbf{C3}).

\begin{sloppypar}
\hi{Our Methodology.} To address these challenges, we propose \oursys, an LLM-powered system for semi-structured document analysis. First, we propose a local-alignment aggregation strategy that transforms OCR-parsed elements into self-contained components, and enriches these components by (1) detecting hierarchical structures of paragraph titles and (2) extracting semantic triples $(title, metadata, data)$ for non-text elements (e.g., tables, charts). Second, we design the \textit{Component-Correlation Tree} (CCTree) to hierarchically organize document components, which can explicitly model inter-component relations and layout-aware distinctions (e.g., separating footnotes and sidebars). We construct CCTree using a bottom-up cascade summarization strategy to propagate concise content upward through the tree. Finally, we employ a question-type-aware retrieval mechanism that supports (1) LLM-guided node selection by titles and metadata, (2) embedding-based fallback to recover missed evidence, and (3) an LLM-based verifier to validate the selected nodes through cross-modal reasoning.
\end{sloppypar}

\hi{Contributions.} 
\marginpar[]{\textbf{\RbColor{R2.O1-(1)}}}
{\RbColor{\llabel{main_contribution} This paper makes the following contributions:}}

\noindent\RbColor{
\noindent(1) We propose \oursys, a novel semi-structured document analysis system that supports diverse document layouts and handles both semantic and location-based questions (see Section~\ref{sec:overview}).}


\noindent\RbColor{
\noindent (2) We design a local-alignment aggregation strategy that transforms OCR-parsed elements into self-contained components, incorporating MLLM-based hierarchy detection for titles and structured semantic extraction for non-text elements. (see Section~\ref{sec:sec:doc-preprocess}).}

\noindent\RbColor{
\noindent (3) We introduce the Component-Correlation Tree (CCTree), a hierarchical representation that captures inter-component relationships and layout-aware distinctions, and enables bottom-up cascade summarization to enhance contextual reasoning (see Section~\ref{sec:sec:cctree}).}

\noindent\RbColor{
\noindent (4) We propose a tree-based data retrieval strategy that integrates question-type-aware retrieval strategy, LLM-guided node selection, embedding-based fallback, and MLLM-based cross-modal verification to support robust evidence localization (see Section~\ref{sec:sec:doc-analysis}).}


\noindent\RbColor{
\noindent (5) Our extensive experiments on standard and specific benchmarks demonstrate significant accuracy improvement (ranging 5.97\% to 61.07\%) over existing methods (see Section~\ref{sec:experiments}).}

\end{sloppypar}

\vspace{-.8em}
\section{Preliminaries}
\label{sec:preliminary}
\vspace{-.1em}

In this section, we introduce the concepts and definition of semi-structured documents (Section~\ref{subsec:docuconcept}), formalize the definition of semi-structured document analysis (Section~\ref{subsec:questiontype}), and discuss the limitations of existing work using a pilot study (Section~\ref{subsec:relatedwork}).

\vspace{-.1em}

\subsection{Semi-Structured Document}
\label{subsec:docuconcept}

\hi{Document Elements.} In this work, we define a \textit{document element}, written as $\bm{\mathit{e}}$, as the atomic unit of a document $\mathcal{D}$, belonging to one of the following five primary categories:


\noindent$\bullet$ \bfit{Text} ($\bm{e^t}$) includes both $(i)$ paragraphs that constitute the main body of the document  (e.g., the two paragraphs about experimental design in Figure~\ref{fig:intro}) and $(ii)$ titles at different hierarchical levels, which represent the theme or summarize the content. For example, in Figure~\ref{fig:intro}, the title ``Experiment ...'' precedes several paragraphs across page 1 and 2, while the title ``Weekly trend of feather score in Exp.1'' leads a chart.


\noindent$\bullet$ \bfit{Tables} ($\bm{e^b}$), including structured tables and semi-structured tables~\cite{semitables}, store attributes and values in cells and reflect relational or nested relations through the spatial arrangement of cells. Tables in semi-structured documents are relatively small in scale but usually contain key records or statistical information for analysis.

\noindent$\bullet$ \bfit{Charts} ($\bm{e^c}$), such as bar charts and line charts, are visual representations of data tables, expressing data relationships through graphical forms (e.g., bars, curves, and sectors), labels, legends, and axes (e.g., scales and grids).

\noindent$\bullet$ \bfit{Images} ($\bm{e^i}$) encompass diverse visual content such as photos, maps, and flow diagrams, representing specific scenes or events. 



\noindent$\bullet$ \bfit{Supplements} ($\bm{e^s}$) include headers, footers, sidebars, and page numbers outside main body, providing auxiliary metadata or navigational cues (e.g., headers with chapter titles, external links). 

\begin{figure}[!t]
    \centering
    \includegraphics[width=\linewidth,  trim=0 0 0 0, clip]{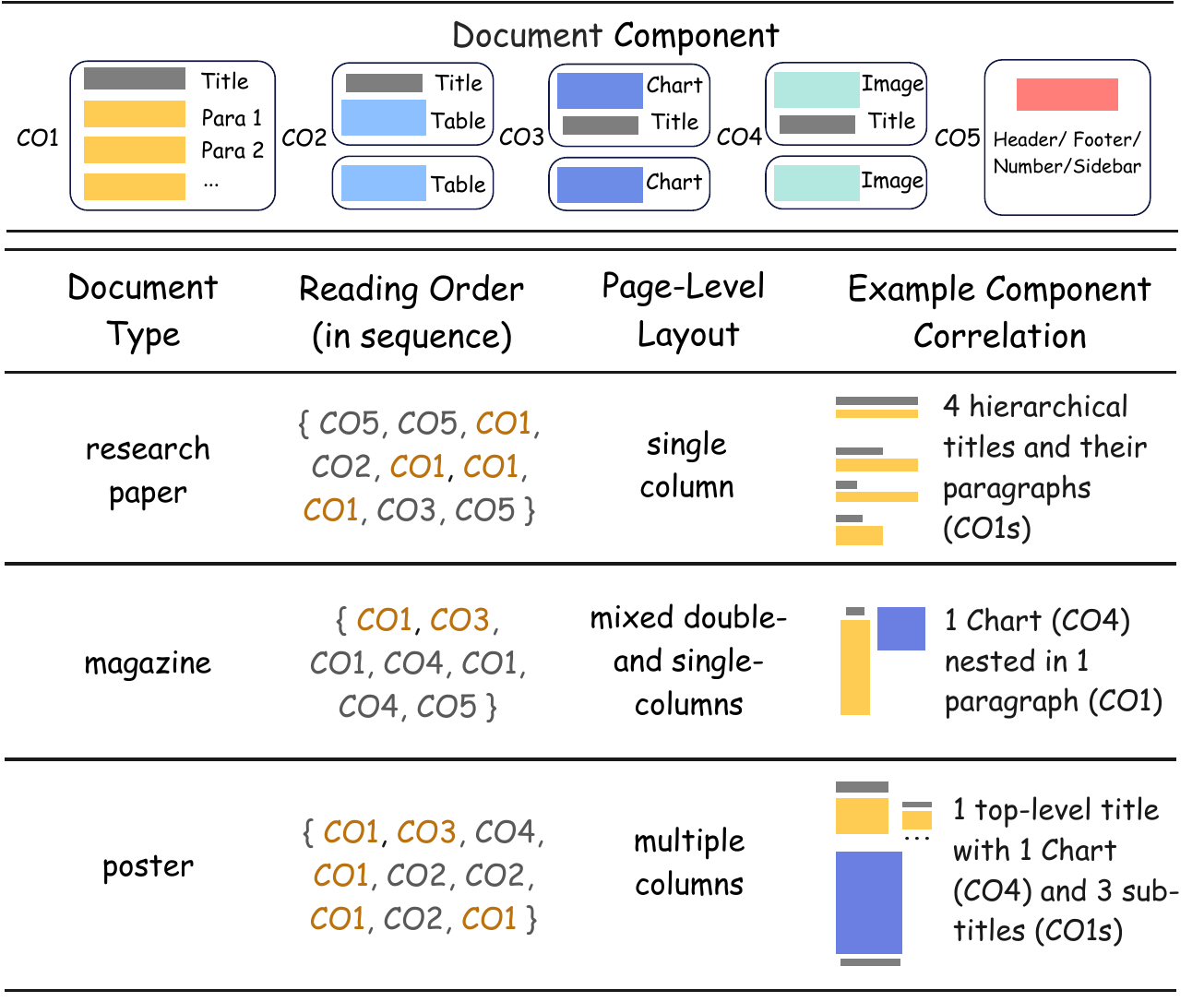}
    \vspace{-2em}
    \caption{Example Semi-structured Document Layouts.}
    \label{fig:multilayout}
    \vspace{-2em}
\end{figure}

\hi{Document Components.} We define a {document component} (\(\bm{\mathit{CO}}\)) as the minimal, self-contained unit of a document, consisting of one or more semantically and visually coherent elements, written as \(\bm{\mathit{CO}} = \{\bm{\mathit{e_{k}^{x}}}\},\ \bm{\mathit{CO}} \subseteq \mathcal{D}\), where \(\bm{\mathit{e_{k}^{x}}}\) denotes an atomic document element. For example, as shown in Figure~\ref{fig:multilayout}, a title followed by multiple paragraphs is grouped as CO1, while a table and its title are aggregated into a table-related component (CO2).

\begin{sloppypar}
\hi{Semi-Structured Document.} A semi-structured document consists of a sequence of layout-ordered components, i.e., $\mathcal{D} = \{\bm{\mathit{CO}_1}, \bm{\mathit{CO}_2}, \dots, \bm{\mathit{CO}_n}\}$, where each $\bm{\mathit{CO}_i}$ is formed by one or more tightly coupled elements, collectively capturing the document’s text and structure information.

\end{sloppypar}

As shown in Figure~\ref{fig:multilayout}, a semi-structured document exhibits all of the following key characteristics: $(1)$ \emph{Diverse element types:} Such documents often contain multiple types of elements within the same page, distinguishing them from pure unstructured text data. $(2)$ \emph{Complex cross-element relationships:} There exist structural and semantic associations among elements, such as hierarchical title structures, spatial grouping, and cross-references between text and tables. $(3)$ \emph{Long document ranges:} The document typically spans a relatively long range (often dozens of pages). 
Unlike long plain text~\cite{longbenchv2}, we focus on challenges in modeling and inference arising from multi-modal elements and complex layouts. 



\subsection{Semi-Structured Document Analysis}
\label{subsec:questiontype}

Given a semi-structured document \(\mathcal{D}\) composed of heterogeneous components (e.g., text, tables, charts) and a natural language question \(\mathcal{Q}\), the goal is to retrieve a subset of components from \(\mathcal{D}\) that are relevant to \(\mathcal{Q}\), such that models can reason over the retrieved components to produce a final answer \(\mathcal{A}\).

\begin{definition}[Semi-Structured Document Analysis]\label{def:doc-analysis} 
Formally, this process is defined as a function \(f: (\mathcal{D}, \mathcal{Q}) \mapsto \mathcal{A}\),  where the answer \(\mathcal{A}\) is derived based on a set of components \(\mathcal{C}_{rel} = \{CO_{k_1}, CO_{k_2}, \dots, CO_{k_m} \mid CO_{k_i} \subseteq \mathcal{D}, CO_{k_i} is\ relevant\ to\ \mathcal{Q}\}\).

\end{definition}




Drawing on existing benchmarks and real-world data (see Figure~\ref{fig:cluster1}), we identify four representative question types in this task.


\noindent\bfit{(1) {Text-Only Questions}} only need the semantics of unstructured text in the document to answer correctly. For example, to answer the question ``\textit{What temperature environment is experiment 3 conducted in?}'', we first retrieve the text component with the title ``\textit{3. Experiment Design and Results}'' , and then perform semantic reasoning over its content to derive the answer ``\textit{0-15 degrees}''.


\noindent\bfit{(2) {Hybrid Data Questions}} need semantic information beyond unstructured text in the document, such as that contained in tables, charts, and images. For example, to answer ``\textit{What is the feather score of Decoquinate group in Exp.3?}'', we first identify the component containing ``\textit{TABLE 1...}'' by semantically interpreting its title and content, and then perform reasoning over the structured data in the table to answer ``\textit{3.20}''.

\noindent\bfit{(3) {Structural Hierarchy Questions}} require understanding the hierarchical structures of the document to answer correctly. For example, to answer {``\textit{From which two aspects does the article analyze the possible causes of the monensin effects?}''}, the system should identify ``\textit{Feather Lesions}''and ``\textit{Changes in Nutritional Requirements}'' as the two parallel chapter titles under the ``\textit{Result Analysis}'', distinguishing them from other textual content. 
\marginpar[]{\textbf{\RbColor{R2.O3-(4)}}}{\RbColor{\llabel{hier_query}
Such questions aim to extract key information or summaries from multiple document chapters~\cite{zhang2025systematic}, rather than to retrieving detailed paragraph content.
}}



\noindent\bfit{(4) {Location-Based Questions}} retrieve components based on specific locations rather than semantics. For example, to answer ``\textit{What is at the bottom-right of page 1?}'', we filter components that meet the criteria based on their location information, including page indices and bounding box coordinates. The result in this case is the figure title ``\textit{Figure 1...}'' and the figure containing two bar plots.




We would like to note that \marginpar[]{\textbf{\RaColor{R1.O1}}}
{\RaColor{\llabel{refine_goals} (1) \oursys centers on structurally modeling document elements and question answering over them. For document parsing that extracts elements from the document, we adopt existing OCR techniques (see Section~\ref{sec:sec:doc-preprocess}), which is not our focus.}}  
\marginpar[]{\textbf{\RbColor{R2.O2}}}{\RbColor{\llabel{complex_query} 
(2) \oursys aims to effectively handle the four representative question types introduced above (see Section~\ref{subsec:overallperformance}), while also supporting more complex multi-hop ones (e.g., involving value aggregations and filtering). Such questions can often be decomposed into combinations of the basic types, for which existing query decomposition techniques~\cite{tang2026straptor} can be readily applied.}}

\begin{figure}[!t]
    \vspace{-.5em}
    \centering
    \hypertarget{fig:precompare}{}
    \includegraphics[width=.95\linewidth,  trim=0 0 0 0, clip]{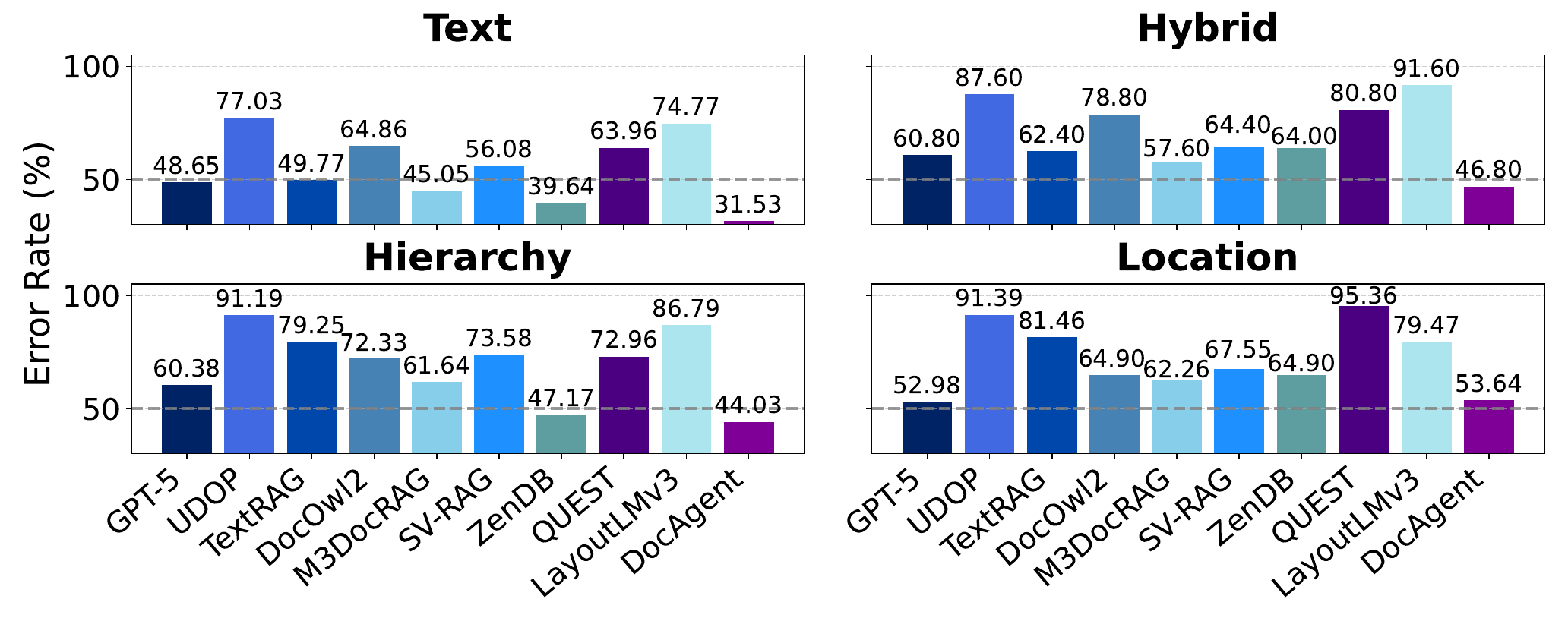}
    \vspace{-1.25em}
    \caption{\RaColor{Error Rate Distribution of Existing Methods.}}
    \label{fig:precompare}
    \vspace{-2em}
\end{figure}

\begin{figure*}[!h]
    \centering
    \includegraphics[width=.98\linewidth,trim={0 0 15 0},clip]{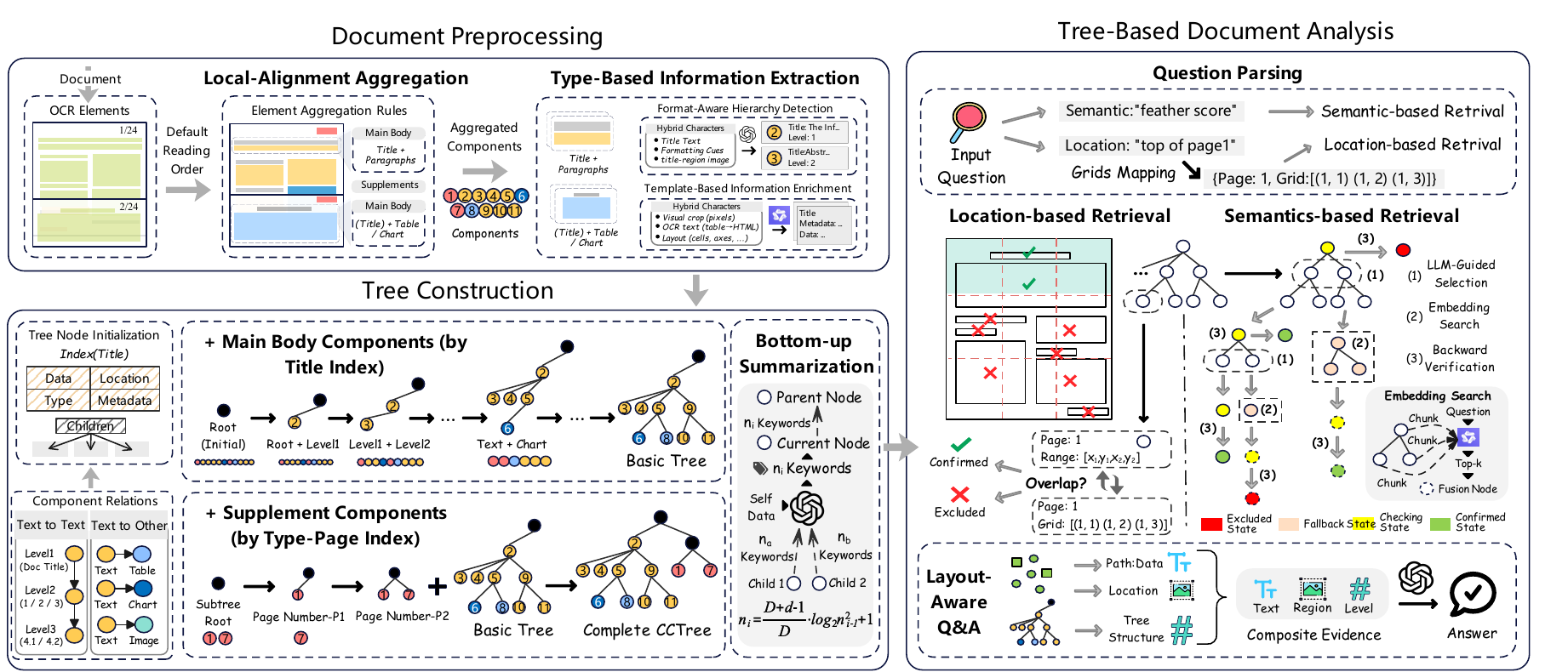}
    \vspace{-.5em}
    \caption{System Overview of \oursys.}
    \vspace{-1em}
    \label{fig:overview}
\end{figure*}

\subsection{Limitations of Existing Methods}
\label{subsec:relatedwork}

Despite recent progress in document analysis, our pilot study (Figure~\ref{fig:precompare}) reveals that existing methods remain inadequate for semi-structured documents, particularly in {\it capturing visual, hierarchical, and layout-specific structures}, which we summarize below.



\hi{Content Extraction Methods.} As discussed in Section~\ref{sec:intro}, methods such as QUEST~\cite{sun2025quest} and EVAPORATE~\cite{arora2023language} transform documents into relational tables to support structured querying. We find QUEST performs poorly on hierarchical and spatial tasks.
The main reason is its failure to preserve structural context during extraction. Moreover, the gap between natural language questions and SQL-style relational queries hinders answer generation.

\hi{Structure Extraction Methods.} 
\marginpar[\textbf{\RbColor{R2.O1-(1)}}]{}{\RbColor{\llabel{more_work5} ZenDB~\cite{zendb} and DocAgent~\cite{sun2025docagent} achieve relatively good accuracy. ZenDB uses visual feature clustering and LLMs to infer partial structures (e.g., separating titles and body text, detecting title levels). However, its error rate of hierarchy analysis approaches 50\%, due to the unreliability of visual feature clustering and weak generalization to non-digital documents. Moreover, its focus on text and titles makes it prone to errors in hybrid data analysis involving tables and charts. In contrast, we aim to natively support multi-modal documents and can uniformly process those without native PDF structures.}}
\marginpar[\textbf{\RbColor{R2.O1-(2)}}]{}{\RbColor{\llabel{more_work4} DocAgent extracts document content into an XML-based tree outline, which is used to guide the agent’s actions and reviews via predefined tools (e.g., keyword search, page and chapter IDs). However, the execution of actions and reviews is only loosely coupled with the XML tree, and the node relationships are largely underutilized. As a result, DocAgent performs poorly on tasks that require structural and layout-aware reasoning, such as hierarchy recognition and location-based analysis, compared to its performance on purely textual tasks.}}

\hi{End-to-End Models.} 
\marginpar[\textbf{\MetaColor{Meta.O1}}]{}
\marginpar[\textbf{\RaColor{R1.O3}}]{}{\RaColor{\llabel{more_work2} End-to-end models, including LayoutLMv3 \cite{huang2022layoutlmv3}, DocFormer~\cite{appalaraju2021docformer}, UDOP~\cite{tang2023unifying}, DocOwl2~\cite{hu2024mplug}, and GPT-5~\cite{openai2025gpt5}), enable end-to-end reasoning through pre-trained models or MLLMs.
LayoutLMv3, DocFormer and UDOP natively support page-level processing with hybrid input of images, OCR-parsed text, and bounding boxes. They employ techniques like discrete multi-modal encoder~\cite{appalaraju2021docformer}, word-patch alignment~\cite{huang2022layoutlmv3} and vision-text-layout unified encoder~\cite{tang2023unifying}. However, limitations in model scale, maximum sequence length (e.g.  512 for text tokenization in LayoutLMv3) and insufficient high-level document structure modeling restrict their performance, particularly on semi-structured documents.
DocOwl2~\cite{hu2024mplug} addresses some of these limitations by incorporating high-resolution visual compression and utilizing a more robust \llm backbone, enabling multi-page input and improved performance, but it still lags behind more advanced MLLMs like GPT-5. Nonetheless, GPT-5 also faces notable challenges in processing semi-structured documents. It often overlooks critical details and struggles to reason over cross-element relationships, such as aligning tables with corresponding paragraphs. Consequently, its performance falls short of structure-aware methods like DocAgent.
}}


\hi{RAG-Based Methods.} Methods such as 
TextRAG~\cite{hegde2023analyzing}, SV-RAG~\cite{chen2024sv}, and M3DocRAG~\cite{cho2024m3docrag} retrieve coarse-grained document segments (e.g., pages or chunks) to mitigate input length limits. We find TextRAG underperforms GPT-5 as it suffers from semantic loss during OCR-based text extraction. M3DocRAG improves performance on text and hybrid tasks by incorporating refined retrieval, but fails to capture hierarchical and locational cues, where its visual encoder lacks the structural grounding needed for these tasks.

\section{System Overview}
\label{sec:overview}

As shown in Figure~\ref{fig:overview}, \oursys is composed of three main modules.

\hi{Document Preprocessing.} To transform raw OCR outputs into self-contained components, we first apply a local-alignment aggregation strategy that groups fragmented elements into layout-aware components. We then extract enriched semantics for two key component types: (1) \textit{Hierarchical titles}, whose hierarchical levels are inferred using a format-aware LLM prompt that incorporates both semantic and visual patterns; and (2) \textit{Non-text elements} (e.g., tables, charts), for which we generate structured semantic triples $(title, metadata, data)$ for information enrichment.


\hi{Tree Construction.} To capture document-level structure beyond linear reading order, we propose the {Component Correlation Tree (CCTree)}, where each document component is represented as a node. We construct CCTree in two stages: (1) \textit{Main content organization}, which groups components based on detected title hierarchies and semantic links between textual and nearby non-text elements (e.g., tables, charts); and (2) \textit{Supplementary content isolation}, which attaches auxiliary components (e.g., sidebars) as separate subtrees to prevent semantic interference. For efficient retrieval, we adopt a bottom-up summarization process that recursively propagates subtree summaries upward, with their length constrained by an empirical formula to balance informativeness and redundancy. 


\hi{Tree-Based Document Analysis.} We design a \emph{question-type-aware retrieval strategy} over the CCTree to accurately locate relevant components. {We first parse locational and semantic cues in the question and adopt different retrieval strategies accordingly:} (1) For \emph{location-based questions}, we select nodes based on spatial metadata, aligning with positional cues in the question. (2) For \emph{semantics-based questions}, we perform a forward search for pruning strategy on CCTree through LLM selection based on node index and metadata, and further apply embedding search as the fallback to recover potential relevant information in unselected subtrees. The MLLM backward verification based on detailed contents and locations is then used to eliminate false positives in LLM selection and embedding search. The final set of evidence along with the CCTree's structure is passed to an MLLM to generate the answer.

\vspace{-.75em}
\section{Document Preprocessing}
\vspace{-.25em}
\label{sec:sec:doc-preprocess}

Document preprocessing aims to transform raw document content into structured and semantically meaningful elements, bridging the gap between low-level visual components (e.g., text boxes, tables, charts) and high-level document structures (e.g., chapters, paragraphs, data-bearing components). 
\marginpar[\textbf{\RaColor{R1.O2}}]{}{\RaColor{\llabel{ocr_use}
In this work, we assume as input a sequence of OCR-parsed elements extracted by tools like PaddleOCR~\cite{cui2025paddleocr30technicalreport}, enabling uniform processing across both digital-native and rasterized documents. Compared to native PDF parsers~\cite{pymupdf,pdfplumber}, OCR offers broader applicability and retains key typographic cues (e.g., font size, indentation) that are essential for document structure modeling. Based on these OCR-parsed elements, we systematically address three main challenges.}}
(1) \textit{Integrating OCR-detected elements into coherent components is nontrivial.} For example, in Figure~\ref{fig:intro}, the title ``Abstract'' and its subsequent paragraphs should be merged into a single component, whereas the title ``Fig. 1: Comparison $\cdots$'' should be combined with its corresponding bar charts.  (2) \textit{Establishing the hierarchical structure of components is essential.} For instance, in Figure~\ref{fig:intro}, the components ``Abstract'' and ``Introduction'' should be placed at the same hierarchical level under the article title.  (3) \textit{Enriching the attributes of non-text components remains challenging.} Tables, charts and images may lack associated textual metadata such as titles or descriptions, and relevant information must be extracted from the visual or structural content of the elements.

To address these challenges, we propose three key techniques: (1) a local-alignment aggregation strategy to detect and organize document components, (2) a format-aware hierarchical relationship detection method to model structural relationships among components, and (3) a template-based information enrichment approach to enrich components containing non-textual elements.



\vspace{-1em}
\subsection{Local-Alignment Element Aggregation}

Direct document analysis at the element level faces challenges in retrieval and reasoning, primarily due to its inability to capture inter-relationships among elements. 
To address this limitation, we introduce the concept of component ($CO=\{e_k^x\}, CO\subseteq \mathcal{D}$) in section~\ref{subsec:docuconcept}.
as a group of {locally} interrelated elements, to facilitate more coherent understanding of document structures. However, the diverse characteristics of elements in semi-structured documents make it challenging to integrate them into components. For instance, adjacent paragraphs under a common title can be concatenated in reading order, but when they span multiple pages, intervening headers or footers may disrupt semantic continuity and cause ambiguity. Moreover, tables, charts and images that appear between paragraphs differ from independent headers and footers, as they are typically titled and linked to the surrounding text.

To tackle the mentioned challenges, documents are first processed by OCR models (e.g., PaddleOCR~\cite{cui2025paddleocr30technicalreport}) to produce sequences of elements arranged in human-preferred reading order. Since the reading order captures the sequential relationships among elements and the element types are identified, it is intuitive to segment the sequence into intervals and organize each subsequence into a specific component. To differentiate various local element layouts, we apply several heuristic rules, as illustrated in Figure~\ref{fig:overview}. Here we introduce several frequently used rules.


\noindent{\textbf{(1) Textual Aggregation:}}
Typically, multiple adjacent paragraphs headed by a title exhibit semantic coherence and thus form a natural component. We aggregate a title and all paragraphs that appear between it and the subsequent title into a component, which can be represented as \(A_1=\{e^t_{title},e^t_{para,1},\dots,e^t_{para,n}\}\).
Each component uses the title as its identifier, concatenates all associated paragraphs as its content, and records the spatial information of its constituent elements as its location, including page indices and bounding box coordinates. 
For example, in Figure~\ref{fig:intro}, the title ``Abstract'' and the paragraphs between it and ``Introduction'' constitute a component.

\noindent{\textbf{(2) Table/Chart Aggregation:}}
For tables, charts and images, their titles (different from those of documents or paragraphs) may appear either before or after them, or in some cases, be absent entirely. For example, if the element immediately preceding a table is identified as a table title and has not yet been assigned to another component, it is designated as the table’s title. Otherwise, if the subsequent element is recognized as a table title, the table adopts it as its title. In the absence of a surrounding title for tables, charts and images, a default title is assigned to them. The aggregated component can be represented by \(A_2=\{[e^t_{title}], e^x\} \cup \{e^x,[e^t_{title}]\}, x\in\{b,c,i\}\), where \([e^t_{title}]\) represents that the title is optional according to different situations mentioned above. The locations of constituent elements are also recorded as part of the component’s attributes.
For example, in Figure~\ref{fig:intro}, the title ``Figure 1: Comparison...'' and the bar plots above it form a component while the title ``Table 1: Effect...'' and the table below it form another component.


\noindent{\textbf{(3) Supplement Element Aggregation:}}
Documents also contain various supplementary elements such as headers, footers, sidebars, and page numbers, which are typically independent on each page. Therefore, each type of supplementary elements on each page is converted into a component and can be represented by \(A_3=\{e^s\}\). The location information of these elements is also recorded within their corresponding component attributes.
For example, the page number ``1/24'' and ``2/24'' in Figure~\ref{fig:intro} each form a component.

\vspace{-.5em}
\subsection{Format-Aware Hierarchy Relation Detection}

Text elements inherently exhibit a nested structure through hierarchical titles, corresponding to the hierarchical organization of the CCTree. However, this structural information is often lost in the linear sequence of elements, and the diverse characteristics of document elements further complicate hierarchy detection. To address this challenge, we propose a format-aware approach for identifying hierarchical relationships among elements.


With the components obtained through local-alignment aggregation, the task of hierarchy detection is mainly about determining the nested levels of paragraphs. Since each component {obtained by rule \(A_1\)} has a title, identifying the level of the title effectively determines that of the entire component.

Intuitively, title levels are determined based on both semantic and textual formatting information. To this end, we extract the list of titles' text and their corresponding locations from the document components to form \(\{e^t_{title,1},\dots,e^t_{title,n}\}\). The recorded locations are used to crop the document regions of each title, which are then concatenated into a unified title sequence vertically. Next we input both the text and image title sequence to an MLLM (e.g., GPT-5) and instruct it to identify the hierarchical level of each title with the input {(i.e., represented by the number of ``\#'' symbols).}

\vspace{-.5em}
\subsection{Template-Based Information Enrichment}

As mentioned above, tables, charts and images can lack explicit titles and are therefore assigned default ones. Moreover, certain information is conveyed in non-textual forms, such as table structures and charts. Due to the diverse structures and rich semantics of these elements, comprehensively extracting their integrated information is challenging. To address this, we propose a template-based method to guide the extraction of such information.



Specifically, we define three types of attributes to facilitate the progressive extraction of integrated information, forming a triplet \(\mathcal{T}(CO)=\big(title(CO), metadata(CO), data(CO)\big), CO\subset A_2\). We prompt MLLM (e.g., Qwen2.5-VL) to generate the triplet in one call. Specifically, the prompt provides differentiated few-shot templates for different primary element. For example, for charts, the prompt instructs the model to focus on axes, labels, and visual trends; whereas for tables, the prompt emphasizes the extraction of schema and representative cell values.  Following this, each component possesses the title, data, and location attributes, along with component-specific attributes reflecting its unique composition.

\section{Tree-Based Document Modeling}
\label{sec:sec:cctree}

\subsection{Component Correlation Tree}
\label{subsec:treedef}

The resulting component sequence does not fully preserve the original document layout (e.g., nested hierarchies or the boundaries between main content and supplementary elements) and thus requires a more structured representation that captures {high-level correlations}. To address this, we propose the Component Correlation Tree (CCTree), which models the relationships between document components while enabling efficient content retrieval.


\hi{Nodes of CCTree.}
As the fundamental unit of the tree, each node \(v\in \mathcal{V}\) in the CCTree represents a document component, retaining the component information and extended with two attributes to facilitate efficient retrieval: $(i)$ node index \(v_{index}\) as a unique identifier and $(ii)$ children \(v_{child}\) recording all child nodes.








\hi{Edges of CCTree.} 
Edges \(\epsilon \in \mathcal{E}\) in the CCTree represent the relationships between nodes in three types.





\noindent{\it (1) \textbf{Text-to-Text Relationship.}}
As the title levels of text paragraphs define the nested hierarchy of a document, the results of format-aware hierarchy detection can be used to establish these relationships. A text component with a higher-level title serves as the parent, whereas components with lower-level titles are designated as children. The relationship can be represented as \(\mathcal{E}_1 = \{(CO_i,CO_j)\mid CO_i,CO_j \subset A_1\}\).

\noindent{\it (2) \textbf{Text-to-Other Relationship.}}
Since non-text components (e.g., tables and charts) typically share a similar semantic theme with adjacent text-based components, we regard them as complementary to the surrounding text. Accordingly, we model the text-based component as the parent node, while the adjacent non-text components are designated as its children. The relationship can be represented as \(\mathcal{E}_2 = \{(CO_i,CO_j)\mid CO_i \subset A_1, CO_j\subset A_2\}\).


\noindent{\it (3) \textbf{Independent Supplementary Relationship.}}
For components containing supplementary elements (e.g., page headers and footers) that are not directly related to the main content, we assign them to an independent branch of the CCTree, using the virtual root node as their parent. The relationship can be represented as \(\mathcal{E}_3 = \{(CO_i,CO_j)\mid CO_i,CO_j \subset A_3\}\).

\vspace{-1em}
\subsection{Two-Stage Tree Construction}

With the defined relationships between tree nodes, we construct the CCTree through a two-stage connection establishment. In the first stage, text-to-text and text-to-other relationships are established, followed by the second stage, which handles the independent supplementary relationships. Algorithm 1 shows the overall process.


\begin{algorithm}[!h]
\caption{CCTree Construction.}
\label{alg:TreeConstruction}

\KwIn{An ordered list $L$ of components}
\KwOut{The CCTree rooted at node $v_r$}
$L_1 \gets [\ ]$\; $L_2 \gets [\ ]$\;
\For{each $CO$ in $L$}{
    \If{$e^s$ in $CO.elements$}{
        $L_2.append(CO)$\;
    }
    \Else{
        $L_1.append(CO)$\;
    }
}
$v_r \gets \mathcal{V}.init()$\;
$v_r.add\_child(Main\_Branch(L_1))$\;
$v_r.add\_child(Supp\_Branch(L_2))$\;
\Return{$v_r$} \;
\end{algorithm}

\hi{Tree Construction with Main Body Components.} The construction for main body components is based on three key factors: the components reading order, their types, and the title levels, as illustrated in Algorithm 2.
It begins by creating a root node with the highest hierarchical level and pushing it onto a stack. Components are then processed sequentially according to the reading order. For a text component, if its title level is lower than that of the node on the stack top, it is linked as the child of the top node and subsequently pushed onto the stack. Otherwise, nodes are continuously popped from the stack until the top node has a higher level than the current component, after which the current component is linked and pushed. For a non-text component, it is directly linked as a child of the top node without being pushed.


Figure~\ref{fig:overview} shows an example. The basic tree of main content is constructed by title index, where the second level title ``\textit{Abstract}'' is linked as the child of first level title ``\textit{The Inf...}'', and the table ``\textit{Table 1...}'' is linked as the child of ``\textit{Experiment ...}''. 


\begin{algorithm}[!t]
\caption{Main Body Branch Construction $Main\_Branch()$.}
\label{alg:SupplementBranchConstruction}

\KwIn{An ordered list $L$ of components}
\KwOut{The root node $r$ of the basic tree}

$v_r \gets \mathcal{V}.init(title\_level=0)$\;
$S \gets stack.init()$ \;
$S.push(v_r)$ \;

\For{each $CO$ in $L$}{
    \If{$CO.elements\ \cap\{e^b,e^c,e^i\} = \varnothing$}{  
        $v \gets \mathcal{V}.init(CO)$ \;
        \While{$S.top().title\_level >= v.title\_level$}{
            $S.pop()$ \;
        }
        $p \gets S.top()$ \;
        $p.add\_child(v)$ \;
        $S.push(v)$ \;
    }
    \Else{  
        $p \gets S.top()$ \;
        $p.add\_child(v)$ \;
    }
}
\Return{$r$} \;
\end{algorithm}

\begin{algorithm}[!t]
\caption{Supplementary Branch Construction $Supp\_Branch()$.}
\label{alg:SupplementaryTreeConstruction}

\KwIn{An ordered list $L$ of components, basic tree root $r$}
\KwOut{The root node $v_r$ of the document tree}
$v_r \gets \mathcal{V}.init(children = [``header", ``footer", ``sidebar", ``number"])$ \;
\For{each $CO$ in $L$}{
    $v \gets \mathcal{V}.init(CO)$\;
    $v_r.children[v.element\_type].add\_child(v)$\;
}
\Return{$v_r$} \;
\end{algorithm}

\hi{Tree Construction with Supplement Components.}  The first stage does not process components of supplementary elements, as they are relatively independent and unsuitable for reading-order-based construction. In the second stage, these components are linked to a separate branch according to their type and page attributes. The detailed procedure is presented in Algorithm 3. 

For the example document in Figure~\ref{fig:intro}, the supplementary branch with page number ``\textit{1/24}'' and ``\textit{2/24}'' is gradually constructed by type-page index. This branch is not correlated with any components from the main content as shown in Figure~\ref{fig:overview}.

\begin{figure}[!t]
    \centering
    \includegraphics[width=1\linewidth]{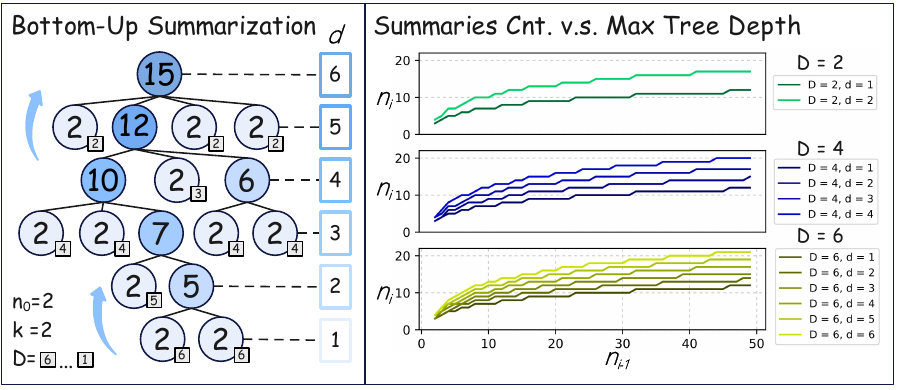}
    \caption{The number distribution of summary keywords in tree nodes with different depth and subtree sizes.}
    \label{fig:summarize}
    \vspace{-1em}
\end{figure}

\subsection{Bottom-up Metadata Summarization.} The document tree construction process connects all components into a complete CCTree. However, the resulting structure is not yet suitable for tree-based retrieval, as higher-level nodes do not inherit information from their lower-level children. Therefore, we propose a bottom-up metadata summarization method that integrates lower-level information into higher-level nodes.



The process starts at the leaf nodes of the CCTree, where each node invokes an LLM to generate several nominal phrases serving as metadata that comprehensively summarize its content. For non-leaf nodes, metadata generation additionally incorporates the metadata of all child nodes. 
Within the same hierarchical level, each node corresponds to a subtree that contains varying amounts of information. During upward information propagation, it is difficult to ensure that all information is completely preserved, leading to inevitable information attenuation. To quantitatively measure and assign the amount of summarized information, we define the summarization as several keywords, and the number of which for each node follows the information attenuation formula:

$$
n_i = \frac{D_i+d_i-1}{D_i} log_2(n_{i-1}^k) + 1
$$

where $D_i$ denotes the maximum depth of the tree that includes node $v_i$, $d_i$ represents the depth of node $v_i$, $n$ indicates the number of summaries of node $v_i$, and $n_{i-1}$ denotes the total number of summaries generated by the children of node $v_i$. The term $\frac{D_i+d_i-1}{D_i}$ serves as the information attenuation coefficient, while $log_2(n_{i-1}^k)$ represents the amount of transmitting information which has a natural decay rate, with $k$ controlling the rate of information growth.

The information attenuation formula determines the number of summaries allocated to each node. After the recursive summarization process, each node retains metadata that encapsulates the semantic content of its entire subtree. The left side of Figure~\ref{fig:summarize} illustrates the bottom-up summarization process with $k=2$, where the number within each node denotes its assigned summaries. The right side compares the number of summaries under different maximum tree depths, revealing the information attenuation effect. For instance, when a parent node aggregates 18 child summaries, it generates 15 summaries after attenuation.

It is worth noting that metadata for non-textual elements have already been generated during the template-based integrated information completion process. To avoid redundant LLM calls and improve efficiency, the existing metadata of these nodes are directly utilized when they serve as leaf nodes.


\section{Tree-Based Document Analysis}
\label{sec:sec:doc-analysis}

In the document tree construction, the entire document is represented as a CCTree, where each node stores textual and locational information. Document analysis involves retrieving relevant evidence nodes and reasoning upon them. The main challenge lies in leveraging the tree structure to optimize retrieval, chasing higher recall and precision. Several retrieval strategies can be applied, each with inherent limitations. Node-by-node traversal guarantees high recall and precision but suffers from high computational cost. Pruning strategies improve efficiency by excluding irrelevant branches but risk missing evidence in pruned subtrees. Embedding-based retrieval offers higher efficiency by directly capturing fine-grained semantics, yet it depends on appropriate chunking and fails to capture long-distance context dependencies.

Considering the strengths and limitations of the above methods, we propose a CCTree retrieval algorithm that integrates MLLM reasoning, pruning strategies, and embedding-based search to balance recall and precision. 


\begin{figure}[!t]
    \centering
    \includegraphics[width=.9\linewidth]{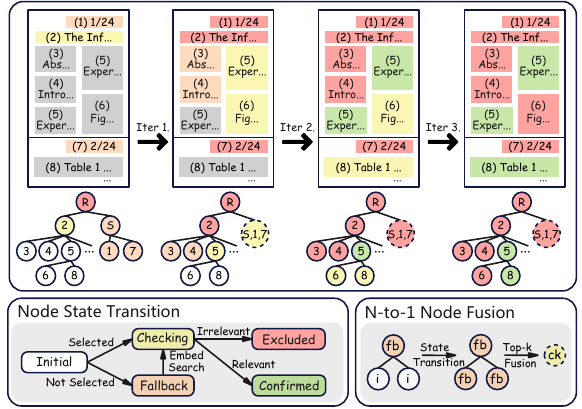}
    \vspace{-.75em}
    \caption{Tree-based Data Retrieval.}
    \vspace{-2.25em}
    \label{fig:tree-retrieval}
\end{figure}

\hi{Semantics- and Location-Aware Retrieval.}
Locational questions requires retrieval based on spatial location rather than semantics. In the CCTree, locations of document elements are stored in the location attributes of nodes, while semantic information is in other attributes. To handle different question types, we adapts location-based retrieval and semantics-based retrieval for locational and semantic cues in the question respectively. Given a document tree, the retrieval algorithm is to evidence nodes in the form of \texttt{(path-data)} pairs and \texttt{(path-location)} pairs. Here, the path represents the sequence of indices from the root to the target node in CCTree, while the data and location are corresponding attributes stored in that node.



\noindent\bfit{(1) Location-Based Retrieval.}
During location-based retrieval, \oursys extracts components whose locations intersect with regions specified in the original question. Specifically, each document page is divided into \(3\times 3 \) grids, and each grid is represented as a tuple $(row, column)$ (e.g., \texttt{(1,1)} corresponds to the top-left, \texttt{(3,2)} to the middle-bottom). Locational cues from the raw question are then mapped to one or more of these grids (e.g., ``bottom of the page'' maps to \texttt{[(3,1),(3,2),(3,3)]}). The CCTree is subsequently traversed, and any node whose location overlaps with the mapped regions is designated as one of the evidence nodes.


\noindent\bfit{(2) Semantics-Based Retrieval.}
During semantics-based retrieval, \oursys performs two iterative steps: $(i)$ forward search executes pruning by deciding whether a node should be added to the candidate evidence node list, which combines a loose LLM selection mechanism with a fallback of embedding search to achieve high retrieval recall, and $(ii)$ backward verification decides whether a node in the candidate list is confirmed as an evidence node or discarded, which leverages MLLM as the verifier to enhance the precision. Figure~\ref{fig:tree-retrieval} shows the retrieval process for the document and question in Figure~\ref{fig:intro}, where different colors represent different node states.



Semantics-based retrieval begins by adding the CCTree root to the candidate evidence node list and initiates an iterative process. In each iteration, forward search and backward verification are performed on each node in the candidate list. The algorithm terminates and returns the results once the candidate list is empty. As shown in the bottom of Figure~\ref{fig:tree-retrieval}, throughout the retrieval process, each node can occupy one of five possible states, including
$(i)$ Initial State ($\mathcal{S}^i$), where the node has not yet been visited,
$(ii)$ Checking State ($\mathcal{S}^{ck}$), where the node has been selected by the forward search or returned by embedding search, and is awaiting backward verification,
$(iii)$ Fallback State ($\mathcal{S}^{fb}$), where the node was not selected by forward search and is scheduled for processing via embedding search.
$(iv)$ Confirmed State ($\mathcal{S}^{cf}$), where the node is determined as relevant by the backward verification as an evidence node.
$(v)$ Excluded State ($\mathcal{S}^{ex}$), where the node is determined as irrelevant by the backward verification to be a non-evidence node.




{\it \textbf{Step 1. Forward Search.}}
For nodes in the candidate evidence list, forward search determines whether each of its child nodes should be added to the list. We prompt the LLM with the question, along with the index and metadata of each child node, to generate a list of indices corresponding to the selected nodes. The selected nodes are those that the LLM deems relevant to the question based on the index and metadata. They are then added to the candidate evidence list, and their state is updated from initial to checking ($\mathcal{S}^i \to \mathcal{S}^{ck}$).

However, since node indices and metadata cannot fully capture the semantics of each node, we employ a fallback of embedding search to capture potential evidence from the sub-trees rooted at nodes not selected during the LLM selection, thereby enhancing recall. As shown in the bottom-right of Figure~\ref{fig:tree-retrieval}, for each node not selected by the LLM, its state and those of all its descendant nodes are updated from initial to fallback ($\mathcal{S}^i \to \mathcal{S}^{fb}$), and the corresponding sub-tree is then flattened, chunked, and embedded for retrieval. The $top\text{-}k$ content pieces most relevant to the question are merged into a fusion node without children, which is subsequently added to the candidate evidence node list ($\mathcal{S}^{fb} \to \mathcal{S}^{ck}$). 



{\it \textbf{Step 2. Backward Verification.}}
The forward search aims to maximize recall by adding as many potential evidence nodes as possible to the candidate list. However, the question may only share certain associations with a node’s index or metadata, or the node may be incorrectly retrieved during the embedding search. To eliminate false positives and ensure retrieval precision, it is therefore necessary to examine the detailed content of each node, which we refer to backward verification.

Specifically, we perform backward verification on each node within the candidate list. An MLLM is prompted to determine whether the node data contains evidence relevant to the question. If the judgment is positive, the node's state is updated to confirmed ($\mathcal{S}^{ck} \to \mathcal{S}^{cf}$) and we output the node. Otherwise, it is just removed from the candidate list and its state is set to excluded ($\mathcal{S}^{ck} \to \mathcal{S}^{ex}$). For an actual CCTree node, both its textual content and the corresponding document region cropped according to its location are provided as inputs to the MLLM. For a fusion node, only its textual content retrieved through the embedding search is used.

\hi{Evidence Aggregation and Answer Generation.}
Finally, we aggregate the results of semantics- and location-aware retrieval into a unified representation to perform final answer generation.
Specifically, three types of evidence are aggregated from the retrieved CCTree nodes. If the retrieval results are empty or deemed irrelevant to the question by the answering model, all nodes in the CCTree will be used to ensure completeness.



\noindent{\it (1) \textbf{Textual Evidence.}}
To enable the MLLM to achieve a fine-grained understanding of the retrieved results, the retrieved CCTree nodes are first converted into \texttt(path-data) pairs, forming a list of textual evidence to be provided to the MLLM as part of its input.


\noindent{\it (2) \textbf{Locational Evidence.}}
Although textual evidence is also extracted for non-textual elements, it fails to capture the structural and formatting information embedded within these elements (e.g., table and chart layouts, font color and size). So we introduce locational evidence to enhance answer generation. Specifically, we crop corresponding document regions with the locations stored in retrieved tree nodes, which is incorporated into the MLLM input.


\noindent{\it (3) \textbf{Hierarchical Information.}}
Both textual and locational evidence provide question-relevant information. However, the global context of the document also plays a crucial role. Therefore, we incorporate the index schema of the CCTree, which represent the document’s hierarchical structure, into the MLLM input as global contextual information.


Using the original question together with the aggregated textual evidence, locational evidence, and hierarchical information, we prompt the MLLM to generate the final answer.

\section{Micro Benchmark Preparation}
\label{ssec:bench}


\marginpar[\textbf{\RaColor{R1.O5}}]{}
{\RaColor{\llabel{dataset_ana}Existing document datasets do not sufficiently cover typical semi-structured document analysis tasks~\cite{tito2023hierarchical,cho2024m3docrag,van2023document}. The documents in MP-DocVQA \cite{tito2023hierarchical} and DUDE~\cite{van2023document} exhibit limited content and structural complexity. While M3DocVQA~\cite{cho2024m3docrag} poses greater structural challenges, its Wikipedia-sourced documents lack element diversity and fail to reflect broader real-world document types. To address these gaps, we introduce a new benchmark \ourbench.}}

\hi{Document Diversity Metrics.} Since basic metrics (e.g., number of pages and elements) fail to capture a document’s layout complexity and element diversity, we introduce two new metrics to characterize structural richness:  Element Diversity ($ED$), which measures the diversity of element types, and Layout Diversity ($LD$), which quantifies the variance in the components' locational distribution.



\noindent{\it \textbf{(1) Element Diversity (ED).}} A document dominated by any single type (e.g., tables, text) tends to exhibit low diversity. Thus we define Element Diversity ($ED$) as the normalized standard variance in the counts of the five element categories (see Section~\ref{sec:preliminary}), i.e.,
\[ ED = \sqrt{\frac{1}{n} \sum_{x}^{} (e^x - \frac{1}{n}\sum_{y}e^y)^2} \Bigg/ \sum_{x}^{}e^x,\ \ where\ \ x,y\in\{t,b,c,i,s\}, \] 
where \(t, b, c, i, s\) denote text, table, chart, image, and supplement respectively and \(n\) is the number of total elements. A larger \(ED\) value indicates lower element diversity.





\noindent{\it \textbf{(2) Layout Diversity (LD).}} 
While $ED$ measures content variety, it does not reflect structural composition. We therefore define \emph{Layout Diversity ($LD$)} to quantify the average number of independent components per page, i.e., \(LD = \frac{N_{\text{comp}}}{N_{\text{page}}}\), where \(N_{\text{comp}}\) denotes the number of independent components and \(N_{\text{page}}\) the total number of pages. 
An independent component refers to a contiguous sequence of elements of the same type; for example, four adjacent tables are counted as one independent component. 
A higher \(LD\) indicates richer layout organization and stronger structural variation.



\hi{Dataset Preparation.} To construct the MMDA benchmark, we begin by sampling 200k candidate documents from over one million real-world documents, spanning academic papers, magazines, financial reports, and presentation slides. Benchmark data is selected from these candidates in three steps. First, to emphasize the layout structures, we annotate each document by overlaying translucent rectangles of maximally distinguishable colors onto regions corresponding to different element types. Second, we compute embeddings using ViT-Base-Patch16-224~\cite{wu2020vit} by the visual features, and apply K-means clustering (with 1,000 clusters, two documents per cluster) to identify 2,000 layout-diverse ones. Third, we rank these documents by their Element Diversity ($ED$) scores and retain the top 537 after filtering out low-quality samples (e.g., documents with blurred or ambiguous content).

\marginpar[]{\textbf{\RcColor{R3.O4}}}{\RcColor{\llabel{qa_curation}
Similar to prior works~\cite{chen2024sv}, we leverage GPT-5~\cite{openai2025gpt5} for generating document QA pairs. We first craft task-specific instructions and few-shot examples, and distinguish between look-up and multi-hop questions. Question types are then sampled by preset ratio (e.g., text:hybrid:hierarchy:location = 9:5:3:3), for each we prompt GPT-5 to generate similar questions following the crafted instructions and examples. The generated QA pairs are reviewed and refined by humans, where annotators prioritize rewriting defective questions into multi-hop queries requiring complex structural and content-based reasoning. All QA pairs are finalized after a second round of correctness verification.}}






\hi{Dataset Statistics.} After data preparation, the MMDA benchmark comprises 1,065 QA pairs across 537 documents. Specifically, the dataset includes 444 text-based questions, 250 hybrid questions involving both text and non-text elements, 159 hierarchy-related questions, 151 location-based questions, and 61 questions targeting formatted text information.

\begin{table}[!t]
\centering
\hypertarget{tab:datasets}{}
\caption{\RaColor{The Statistics of Document Datasets.}}
\label{tab:datasets}
\vspace{-1em}
\resizebox{.95\linewidth}{!}{
{
\begin{tabular}{lccccccc}
\toprule
\textbf{Bench} & \textbf{DUDE} & \textbf{MP-DocVQA} & \textbf{M3DocVQA} & \textbf{\ourbench} \\
\midrule
ED & 0.313 & \underline{0.305} & 0.323 & \textbf{0.264}\\
LD & \underline{1.35} & 0.94 & 1.26 & \textbf{1.85}\\
Nodes & 17.2 & 27.6 & \underline{38.3} & \textbf{49.2}\\
Leaves & 12.6 & 20.8 & \underline{26.2} & \textbf{37.7}\\
Levels & 3.36 & 3.66 & \textbf{4.96} & \underline{4.23}\\
\bottomrule
\end{tabular}}}
\vspace{-1em}
\end{table}

\begin{figure}[!t]
    \centering
    \hypertarget{fig:cluster1}{}
    \includegraphics[trim={0.5cm 0cm 3.5cm 0cm}, clip, width=1\linewidth]{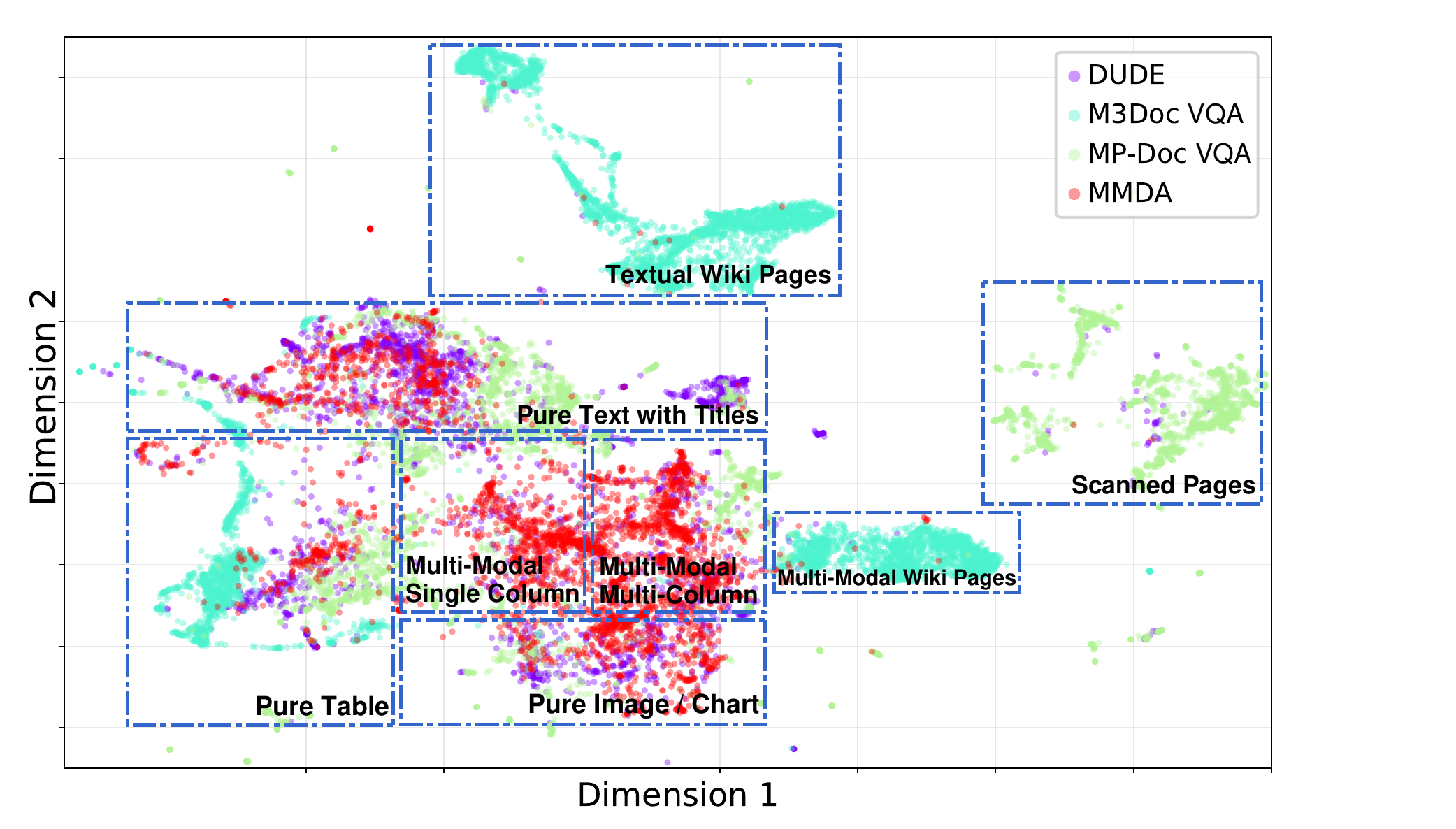}
    \vspace{-2.25em}
    \caption{\RaColor{Distribution of Document Layout Characteristics.}} 
    \vspace{-1.5em}
    \label{fig:cluster1}
\end{figure}

\hi{Dataset Comparison.} \marginpar[\textbf{\RaColor{R1.O5}}]{} {\RaColor{\llabel{cluster} Figure~\ref{fig:cluster1} compares the document layout distribution of DUDE, MP-DocVQA, M3DocVQA and \ourbench in the embedding space. While all datasets cover basic layouts (e.g., pure text, table, image or chart), \ourbench demonstrates broader coverage of complex single-/multi-column structures with multiple modality (e.g., mixed text and tables on a page). M3DocVQA and MP-DocVQA own some distinct clusters, as their Wikipedia pages and scanned pages exhibit unique styles. Other datasets include a limited number of similar document types, contributing to the distinct clusters.}}


\marginpar[\textbf{\RaColor{R1.O5}}]{}
\marginpar[\textbf{\RcColor{R3.O3}}]{}{\RaColor{\llabel{dataset_cmp} Besides, we compute several dataset statistics, including Element Diversity (ED), Layout Diversity (LD), and the average number of nodes, leaves, and levels in the CCTree for each dataset. As shown in Table~\ref{tab:datasets}, \ourbench exhibits lower ED (0.264 vs. 0.305 for MP-DocVQA), higher LD (1.85 vs. 1.35 for DUDE), and more tree nodes (49.2 vs. 38.3 for M3DocVQA), reflecting its greater element variety, more complex layout distributions, and richer content. While \ourbench has slightly fewer CCTree levels than M3DocVQA, this is partly due to M3DocVQA's focus on Wikipedia documents, which feature inherently deep hierarchical structures.}} 

\hi{Evaluation Metrics.} 
\marginpar[\textbf{\RbColor{R2.O3-(4)}}]{}{\RbColor{\llabel{open_end} To accurately evaluate diverse real-world questions (e.g., open-ended ones~\cite{zhang2025systematic}), we adopt two main metrics: 
(1) \textbf{ACNLS}: We extend the standard ANLS~\cite{biten2019scene}, which uses edit distance to tolerate minor textual variations. Since verbose responses may incur low ANLS scores despite containing correct answers, we assign a score of 100\% when the response contains the reference answer. 
(2) \textbf{AIC-Acc}: We employ LLM-based judgment to handle semantic equivalence across varied phrasings, assigning 1 for correct answers and 0 otherwise. Similarly, responses containing the reference answer receive 100\% to accommodate verbosity.}}

\vspace{-.25em}
\section{Experiments}
\label{sec:experiments}

\begin{figure*}[!h]
    \centering
    \hypertarget{fig:trend}{}
    \includegraphics[trim={0.1cm 0.5cm 1cm 0.5cm}, clip, width=.9\linewidth]{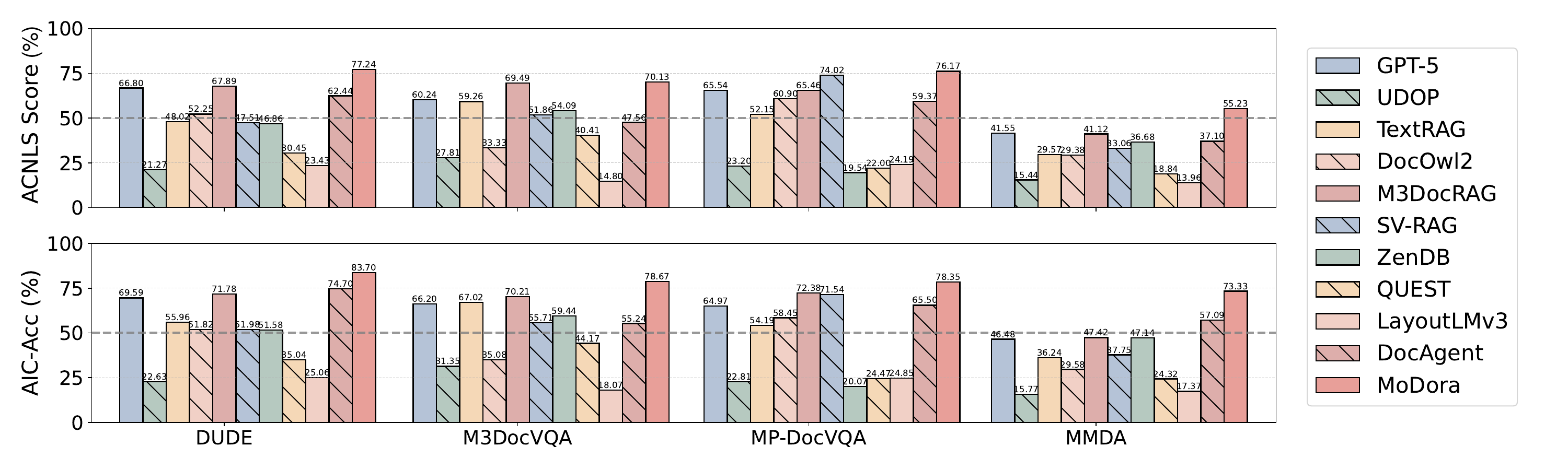}
    \caption{\RaColor{Performance Comparison Between Different Baselines on Four Benchmarks.}}
    \label{fig:trend}
    \vspace{-1em}
\end{figure*}

\subsection{Experimental Setup}
\label{subsec:setup}


\hi{Models.} 
\marginpar[]{\textbf{\RcColor{R3.O5}}}{\RcColor{\llabel{model_setup}
We employ PaddleOCR~\cite{cui2025paddleocr30technicalreport} for document elements parsing and Qwen3-Embedding-8B~\cite{qwen3embedding} for embedding search. For relatively simpler tasks such as information enrichment, summarization generation, forward search and backward verification, we locally implement Qwen2.5-VL-7B-Instruct~\cite{qwen2.5-VL}. GPT-5 is applied for more challenging tasks, including hierarchy detection, question parsing and final question answering in \oursys, as well as for the uniformly external LLM-API used in baselines.
}}


\begin{sloppypar}
\hi{Baselines.} We compare eight typical baselines including UDOP~\cite{tang2023unifying}, DocOWL2~\cite{hu2024mplug}, M3DocRAG~\cite{cho2024m3docrag}, SV-RAG~\cite{chen2024sv}, TextRAG~\cite{hegde2023analyzing}, QUEST~\cite{sun2025quest}, ZenDB~\cite{zendb} and GPT-5~\cite{openai2025gpt5}. 
\end{sloppypar}

\hi{Benchmarks.} We evaluate \oursys and aforementioned baselines on the DUDE and \ourbench, with moderate document length and diverse document layouts (see Figure~\ref{fig:cluster1}). We separately measure their performances using ACNLS and AIC-Acc (see Section~\ref{ssec:bench}). 

\noindent \textbf{Implementation}. We conduct experiments on a workstation with 2 Intel Xeon Platinum 8352V CPU (2.10 GHz), 1 TB RAM, and 8 NVIDIA RTX 4090 GPUs.
\marginpar[]{\textbf{\RbColor{R2.O3-(2)}}}{\RbColor{\llabel{parameter} 
We set $n_0=2$ and $k=2$ for bottom-up summarization, $temperature = 1$ for LLM generation, and retrieve 3 chunks or pages via embedding search. Other baseline parameters follow the default settings in their open-source repositories~\cite{sht, docagent, m3raggit}.
}}

\subsection{Performance Comparison}
\label{subsec:overallperformance}

\hi{Overall Performance.}
\marginpar[]{\textbf{\MetaColor{Meta.O2}}}\marginpar[]{\textbf{\RaColor{R1.O4}}} \marginpar[]{\textbf{\RaColor{R1.O5}}} \marginpar[]{\textbf{\RbColor{R2.O1-(1),(2)}}}{\RaColor{\llabel{main_discussion} 
As shown in Figure~\ref{fig:trend}. \oursys outperforms all baselines across all datasets and metrics, achieving 9.00\% higher AIC-Acc than the best-performing baseline on DUDE, 8.46\% on M3DocVQA, 5.97\% on MP-DocVQA, and 16.24\% on \ourbench. Its superior performance can be attributed to three key factors.}} 

\RaColor{First, \oursys captures both semantic and structural information from semi-structured documents, encoding them into unified component representations (Section~\ref{sec:sec:doc-preprocess}). This design benefits the subsequent tree construction and analysis, enabling them to operate at a fine-grained yet complete level. In contrast, baselines do not involve this step: some (e.g., GPT-5, DocOwl2, UDOP, LayoutLMv3) process the entire document as a whole, some (e.g., SV-RAG, M3DocRAG) segment by full pages without capturing intra-page structures, and others (e.g., TextRAG, ZenDB) rely solely on textual content, ignoring layout and visual cues, limiting their ability to align and reason over structure and content of elements.}



\RaColor{Second, \oursys leverages node indexing and summarization within the CCTree to enable hierarchical organization and multi-level abstraction, facilitating accurate and efficient retrieval from documents with multi-level nested structures. In contrast, baselines struggle with structured retrieval and reasoning. For instance, TextRAG becomes overwhelmed by excessive text due to the lack of guidance from document titles. For ZenDB, it only models titles and text while overlooking other critical elements like images and charts. Additionally, DocAgent and ZenDB ignore the fine-grained location (bounding box) of elements and fail to distinguish between main content and supplements in their hierarchical structure, leading to inferior performance compared to \oursys.}

\RaColor{Third, the question-type-aware retrieval strategy in \oursys dynamically selects appropriate retrieval method based on question types, fully leveraging both the locational and semantic information encoded in the tree. For semantic retrieval, the LLM-guided selector, embedding search, and MLLM verifier collaborate to balance precision and recall, effectively handling complex evidence distributions within documents (Section~\ref{sec:sec:doc-analysis}). In contrast, embedding-based methods are either too narrow, capturing only local semantics within chunks (e.g., TextRAG), or too coarse, retrieving entire pages without fine-grained discrimination (e.g., M3DocRAG and SV-RAG).}

\RaColor{Interestingly, we observed that both SV‑RAG and M3DocRAG outperform other baselines on MP‑DocVQA. This is attributed to their page retrieval modules (Col‑retrieval~\cite{chen2024sv} for SV‑RAG and ColPali~\cite{faysse2025colpali} for M3DocRAG) being trained on DocVQA question‑page pairs, which essentially correspond to the page‑retrieval task in MP‑DocVQA. In contrast, \oursys, which extracts content via OCR and then performs document modeling and analysis, achieves superior performance without specific training.} 

\hi{Evaluation Across Question Types.} 
{\RaColor{\llabel{type_discussion} As shown in Table~\ref{tab:type}, \oursys performs best across all four typical question types.\\
\noindent{\it (1) \textbf{Hierarchy Questions.}} \oursys achieves the highest accuracy on hierarchy-based questions, demonstrating the CCTree’s effectiveness in capturing document structure. By explicitly organizing elements based on heuristic aggregation rules and modeling component hierarchies, \oursys outperforms all baselines. While DocAgent and ZenDB rank second and third on this type, benefiting from their structural designs (XML and SHT trees), they fall significantly behind \oursys due to limited robustness in handling complex layouts during tree construction and retrieval.\\
\noindent{\it (2) \textbf{Textual Questions.}} For paragraph text questions, \oursys again outperforms all competitors. Its component-aware design ensures fine-grained representation of relevant textual blocks while avoiding unnecessary noise. DocAgent follows in performance, as its agent framework enables \llm to engage in thorough reasoning during retrieval and answering. However, its keyword-matching for semantic-based retrieval fails to fully exploit the tree structure, and does not incorporate summarization to condense detailed document content. Consequently, it still lags behind \oursys.\\
\noindent{\it (3) \textbf{Hybrid Questions.}} While \oursys maintains the top performance, hybrid questions pose a greater challenge due to the need to jointly interpret textual content alongside tables, charts and images. 
DocAgent ranks second because it includes tables and images in its XML tree structure outline, and allows its agent to retrieve them by table IDs or image IDs. But this approach sometimes fails due to the lack of matching IDs for the corresponding element or poor summarization of the content.\\
\noindent{\it (4) \textbf{Location Questions.}} For questions requiring location-based reasoning, \oursys again leads, leveraging locations in the CCTree and grid-based mapping technique to pinpoint relevant elements. GPT-5 ranks second, as its powerful visual comprehension capabilities allow it to analyze page layouts directly from full-document input.
DocAgent has comparable performance to GPT-5 because its agent can retrieve document pages by number. However, it is not designed to consider the specific location (bounding box) of elements on a page, leading to failures in analysis requiring finer-grained localization, similar to using GPT-5 directly.
}}

\begin{table}[!t]
\centering
\hypertarget{tab:type}{}
\caption{\RaColor{AIC-Acc Performance by Question Types.}}
\label{tab:type}
\vspace{-1em}
\setlength{\tabcolsep}{3pt}
\begin{tabular}{c c c c c c c c}
\toprule
\textbf{Method} & \textbf{Hierarchy}  & \textbf{Text} & \textbf{Hybrid} & \textbf{Location} \\
\midrule
GPT-5	& 39.62	& 51.35	& 39.20	& \underline{47.02} \\
UDOP	& 8.81	& 22.97	& 12.40 & 8.61 	\\
TextRAG	& 20.75	& 50.23	& 37.60	& 18.54 \\
DocOwl2	& 27.67	& 35.14	& 21.20	& 35.10 \\
M3DocRAG	& 38.36	& 54.95	& 42.40	& 37.75 \\
SV-RAG	& 26.42	& 43.92	& 35.60	& 32.45 \\
ZenDB	& 52.83	& 60.36	& 36.00	& 35.10 \\
QUEST   & 27.04 & 36.04 & 19.20 & 4.64 \\
\RaColor{LayoutLMv3}   & \RaColor{13.21} & \RaColor{25.23} & \RaColor{8.40} & \RaColor{20.53} \\
\RaColor{DocAgent}   & \RaColor{\underline{55.97}} & \RaColor{\underline{68.47}} & \RaColor{\underline{53.20}} & \RaColor{46.36} \\
\oursys	& \textbf{76.73}	& \textbf{79.95}	& \textbf{68.00}	& \textbf{68.21} \\
\bottomrule
\end{tabular}
\vspace{-1.5em}
\end{table}


\subsection{Fine-Grained Result Analysis}
\label{subsec:finegrained}

\hi{Analysis of Tree Construction.}
A well-constructed CCTree, built upon component representations, should accurately and coherently capture the content and structure of a semi-structured document. Manual observation finds the current construction method is applicable to most documents, while imperfect tree construction accounts for only about 20\% of the observed analysis errors.

Suboptimal tree structures arise from OCR errors that misclassify titles as plain text, and LLM deviations in hierarchy detection. Even when certain nodes are merged or misplaced due to hierarchy inaccuracies, their metadata can still be propagated upward through the bottom-up metadata generation process and effectively utilized during tree-based retrieval, which demonstrates the robustness of the CCTree. \marginpar[\textbf{\RcColor{R3.O5}}]{}{\RcColor{\llabel{vl_error} In addition to structural errors, information enrichment may also generate incorrect descriptions of images, tables, charts, and propagate such errors into the content of tree nodes. We find that such errors occur in only $\sim$3\% of the generated results.}} 

\hi{Analysis of Tree-based Retrieval.}
Tree-based retrieval process aims to identify relevant evidence nodes within the tree, balancing high precision and high recall to achieve more accurate analysis. Through manual inspection of several retrieval cases, we observe that during location-based retrieval, \oursys effectively extracts page and region evidence from the question to perform region mapping, thereby achieving high recall. 

In contrast, semantics-based retrieval using the LLM-based selector and embedding search effectively identifies relevant nodes, while the MLLM verifier balances precision and recall. Manual observation reveals that the selector and embedding search recall all required evidence with a rate of 91\%, and the verifier filters 84\% of irrelevant nodes, though it excludes 12\% of necessary evidence.


We also find that the inclusion of a few irrelevant nodes usually does not affect obtaining correct answer, whereas errors can still arise despite perfect evidence nodes (limited by analysis complexity and reasoning capability of the MLLM). 

\hi{Different Backbone LLMs.} To assess the impact of different backbone LLMs on \oursys, we replace GPT-5 with several mainstream LLMs for analysis, as shown in Table~\ref{tab:basic}. Although ACNLS and AIC-Acc fluctuate within a certain range, the overall performance remains significantly higher than that of all baselines. These results demonstrate that the superior performance of \oursys stems from its inherent effectiveness of design rather than reliance on any particular LLM. Consequently, users can flexibly adopt different LLMs as the backbone of our framework according to their preferences and requirements.

\begin{table}[!t]
\centering
\hypertarget{tab:basic}{}
\caption{Performance under Different LLM Backbones.}
\label{tab:basic}
\vspace{-1em}
\begin{tabular}{lcc}
\toprule
\textbf{LLM} & \textbf{ACNLS} & \textbf{AIC-Acc (\%)} \\
\midrule
GPT-5 & 55.23 & 73.33 \\
GPT-4o & 57.12 & 70.05 \\
Gemini-2.5-Flash & 55.58 & 71.27 \\
Qwen3-Max & 48.02 & 62.63 \\
\bottomrule
\end{tabular}
\vspace{-1em}
\end{table}

\begin{table}[!t]
\centering
\hypertarget{tab:sht}{}
\caption{\RbColor{Performance Comparison using datasets from~\cite{sht}.}} 
\label{tab:sht}
\vspace{-.75em}
\setlength{\tabcolsep}{3pt}
\begin{tabular}{c c c c c c c c}
\midrule
\textbf{Dataset} & \multicolumn{3}{c}{\textbf{Civic}} & \textbf{Qasper} & \textbf{Contract}\\ \hline
\textbf{Metric} & \textbf{accuracy}  & \textbf{precision} & \textbf{recall} & \textbf{f1 score} & \textbf{accuracy} \\ \hline
ZenDB	& 71.8	& 81.5 & 42.7 & 31.8 & 85.7\\
\oursys	& 77.3	& 89.4	& 71.1	& 35.6 & 90.6\\
\bottomrule
\end{tabular}
\vspace{-1em}
\end{table}

\hi{Analysis of the ACNLS metric.}
\marginpar[\textbf{\RbColor{R2.O3-(1)}}]{}{\RbColor{\llabel{whole_output}
 To verify the effectiveness of ACNLS metric, we conduct a test by using the whole-document content as the answer. It results in 43.85 on \ourbench and 64.72 on DUDE. These scores are much lower than the containment score 100, and also lower than the score of \oursys shown in Figure~\ref{fig:trend}. This demonstrates that the advantage of \oursys stems from the retrieval and reasoning over document content, rather than from a special metric design or the whole-document extraction.}} 
 
\hi{Comparisons using ZenDB benchmarks.}
 \marginpar[]{\textbf{\RbColor{R2.O1-(1)}}}{\RbColor{\llabel{sht_test} We further evaluate \oursys against ZenDB~\cite{sht} using the same datasets and evaluation metrics reported in~\cite{sht}. As shown in Table~\ref{tab:sht}, \oursys consistently delivers substantially stronger performance, demonstrating robust generalizability across seven benchmarks (Civic, Qasper, Contract, DUDE, MMDA, M3DocVQA, and MP-DocVQA) and a diverse set of evaluation metrics (precision, recall, f1, ACNLS, AIC-CC).}}

\begin{table}[!t]
\centering
\caption{\RcColor{LLM-API Based Token Consumption on MMDA.}}
\label{tab:cost}
\vspace{-1em}
\setlength{\tabcolsep}{3pt}
\begin{tabular}{lccccc}
\hline
\multirow{2}{*}{\textbf{Tokens}} & \multicolumn{3}{c}{\textbf{MoDora}} & \multirow{2}{*}{\textbf{DocAgent}} & \multirow{2}{*}{\textbf{GPT-5}} \\
 & Preprocess & Analysis & SUM &  &  \\
\hline
Prompt     & 369k  & 6654k & 7023k & 52773k & 771k \\
Completion & 1084k & 604k  & 1688k & 2523k  & 876k \\
\hline
\end{tabular}
\end{table}

\begin{figure*}[!t]
    \centering
    \includegraphics[width=1\linewidth,trim={0 5em 0 0},clip]{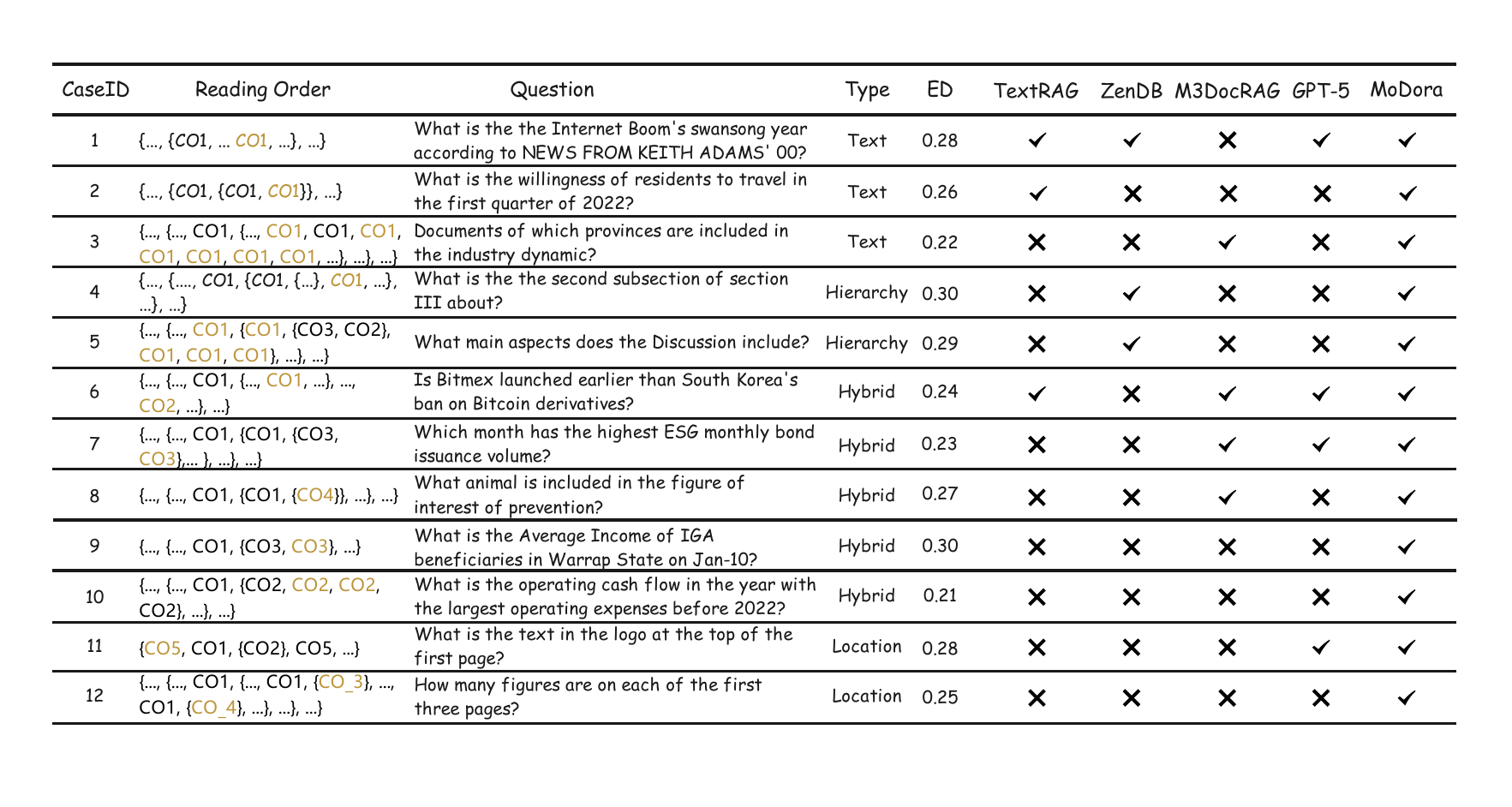}
    \vspace{-1.5em}
    \caption{Case Study on \ourbench Benchmark.}
    \label{case_study}
    \vspace{-.75em}
\end{figure*}

\subsection{API-Based LLM Cost Analysis}
\label{subsec:cost}
\marginpar[]{\textbf{\RcColor{R3.O1}}}{\RcColor{\llabel{cost}Table~\ref{tab:cost} reports the API cost of \oursys and two baselines (DocAgent, GPT-5) on \ourbench. With local-model-offloading optimization (Section~\ref{subsec:setup}), \oursys consumes 7,023k prompt tokens and 1,688k completion tokens in total, averaging \$0.025 per query. The pipeline requires 1, 0, and 4 API calls per query for preprocessing, tree construction, and analysis, respectively. By contrast, DocAgent incurs substantially higher costs (52,773k prompt tokens, 2,523k completion tokens; \$0.09 per query) due to its multi-round interactions. Directly prompting GPT-5 with the full document is the cheapest option (771k prompt tokens, 876k completion tokens; \$0.01 per query), but sacrifices accuracy. 
Thus, \oursys strikes a practical balance: its per-query cost remains comparable to simpler baselines while delivering over 15\% higher accuracy on \ourbench (Figure~\ref{fig:trend}).}}

\begin{table}[!t]
\centering
\hypertarget{tab:ablation}{}
\vspace{-.75em}
\caption{\RaColor{Ablation Study Performance (AIC-Acc).}}
\label{tab:ablation}
\vspace{-.75em}
\begin{tabular}{lcc}
\toprule
\textbf{Methods} & \textbf{DUDE} & \textbf{MMDA Bench} \\
\midrule
Full Model & 83.70 & 73.33\\
w/o Textual Evidence & 70.32(-13.38) & 60.56(-12.77)\\
w/o Locational Evidence & 71.29(-12.41) & 68.26(-5.07)\\
w/o Forward Search & 78.35(-5.30) & 70.89(-2.44)\\
\RaColor{w/o Tree Structure} & \RaColor{68.37(-15.33)} & \RaColor{57.84(-15.49)}\\
\RaColor{w/o Component Construction} & \RaColor{55.96(-27.74)} & \RaColor{36.24(-37.09)}\\
\bottomrule
\end{tabular}
\vspace{-1.25em}
\end{table}

\subsection{Ablation Study}
\label{subsec:ablation}

We conduct an ablation study on four design variants of \oursys, and the results are presented in Table~\ref{tab:ablation}.

\hi{Without Textual Evidence.} We remove textual information from the inputs to the MLLM during node verification and answer generation, leading to substantial AIC-Acc drops on DUDE (-13.38\%) and \ourbench (-12.77\%). This highlights MLLMs' limitations in extracting key information and reasoning solely from document page images, underscoring the critical role of textual information in providing essential contextual cues.

\hi{Without Locational Evidence.} We remove locational document region evidence from the MLLM inputs to assess the expressiveness of pure text. This results in performance drops of 12.41\% on DUDE and 5.07\% on \ourbench, highlighting the importance of incorporating original document regions. They mitigate information loss during information enrichment for non-textual elements, and provide formatting cues (e.g., font size, style, color) absent in plain text.

\hi{Without Forward Search.} 
\marginpar[\textbf{\RaColor{R1.O4}}]{}\marginpar[\textbf{\RbColor{R2.O3-(3)}}]{}{\RaColor{\llabel{re_ablation} We disable the LLM-based forward search in \oursys, relying solely on embedding search and backward verification on the tree. 
This leads to AIC-Acc drops of 5.3\% on DUDE and 2.44\% on \ourbench, highlighting that retrieval guided by titles and metadata significantly improves performance.}} 

\hi{Without Tree Structure.} \marginpar[\textbf{\RaColor{R1.O4}}]{} \marginpar[\textbf{\MetaColor{Meta.O2}}]{} \marginpar[\textbf{\RcColor{R3.O2}}]{}{\RaColor{\llabel{hi_ablation} We evaluate the impact of hierarchical representation by performing question answering on flattened components using the same retrieval strategies. Without the tree structure, AIC-Acc drops significantly by -15.33\% and -15.49\% on DUDE and \oursys respectively, demonstrating the effectiveness of CCTree's structural modeling.}}

\hi{TextRAG.} 
\marginpar[\textbf{\RaColor{R1.O4}}]{}
\RaColor{\llabel{all_ablation} This baseline represents \oursys without three main components (i.e., {component reconstruction, hierarchical representation, retrieval strategy}), resulting in the most severe performance degradation, surpassing the impact of removing the hierarchical representation or retrieval strategy individually.}


\subsection{Case Study}
\label{sec:sec:case-study}
In Figure~\ref{case_study}, for questions in different types, we select representative cases to compare the performance of various methods. The reading order sequence of  Doc Layout illustrates the tree structure around the evidence nodes (highlighted in gold). Based on the reading order sequence in Figure~\ref{fig:multilayout}, the nested structure in the tree is denoted by "\{ \}", and evidence nodes are highlighted in gold. 

We observe three main findings from the case study:
(1) GPT-5 shows limited capability in information awareness and reasoning over semi-structured documents. In Cases 3–5, it retrieves only partial sets of provinces and aspects compared to the ground truth; in Case 10, it fails to integrate information from two tables for correct reasoning.
(2) Embedding-based retrieval struggles to capture document hierarchies. Although TextRAG (Cases 1, 2, 6) and M3DocRAG (Cases 6–8) perform reasonably on retrieving isolated values or pages, they fail to preserve hierarchical structures critical for Cases 4 and 5, often returning irrelevant content.
(3) Text-only representations result in significant information loss for non-textual elements such as tables, charts, and images. As a result, methods like TextRAG and ZenDB, which convert documents into pure text, are unable to answer most hybrid questions in Cases 6–10.




\vspace{-.75em}
\section{Conclusion}
\label{sec:conclusion}

In this paper, we proposed \oursys, an LLM-powered framework for semi-structured document analysis. We adopted a local-alignment aggregation strategy to convert OCR-parsed elements into layout-aware components and extract type-specific information from hierarchical and non-text elements. Then we designed a Component-Correlation Tree (CCTree) to hierarchically organize components and capture inter-component relations through bottom-up summarization. We also proposed a question-type-aware retrieval strategy combining layout-based grid partitioning and LLM-guided semantic pruning. Experimental results on public datasets and our MMDA benchmark showed that \oursys significantly outperformed state-of-the-art methods.


\marginpar[]{\textbf{\RaColor{R1.O2}}}{\RaColor{\llabel{native_use}
In the future, we plan to enhance \oursys in three main aspects. First, we will integrate native PDF structure and typographic cues not preserved by current OCR techniques to enrich the extracted information of semi-structured documents. Second, we aim to extend the CCTree to a cross-document forest for multi-document QA. Third, we will explore more advanced retrieval strategies with dynamic traversal, such as backtracking and query-driven skipping, to more effectively address potentially complex queries.
}}

\clearpage
\newpage
\bibliographystyle{ACM-Reference-Format}
\bibliography{sample-clean}

\end{document}